\documentclass[twocolumn,floatfix,linenubers]{aastex62}
\pdfoutput=1 
\usepackage{amsmath,amstext}
\usepackage[T1]{fontenc}
\usepackage[utf8]{inputenc}
\usepackage{apjfonts} 
\usepackage[figure,figure*]{hypcap}
\usepackage{amsmath,amssymb,amsfonts,graphicx}
\usepackage{color}
\usepackage{graphicx}
\usepackage{tabularx}
\usepackage{footnote}
\usepackage{subfigure}

\shorttitle{PHOTOMETRY AND SPECTROSCOPY OF CONTACT BINARIES}
\shortauthors{Panchal et al.}

\received{July XX, 2020}

\accepted{July XX, 2020}

\submitjournal{AJ}

\begin{document} 

\title{Exploring Contact Binaries: Observational Analysis of four W Uma Binaries Using Photometry and Spectroscopy}

\author{Panchal, A.}
\affiliation{Physical Research Laboratory (PRL), Ahmedabad, Gujarat, India.}
\email{alaxenderpanchal@gmail.com}

\author{Joshi, Y. C.}
\affiliation{Aryabhatta Research Institute of observational sciencES (ARIES), Nainital, Uttrakhand, India.}

\author{Joshi, V.}
\affiliation{Physical Research Laboratory (PRL), Ahmedabad, Gujarat, India.}

\author{Chakraborty, A.}
\affiliation{Physical Research Laboratory (PRL), Ahmedabad, Gujarat, India.}

\begin{abstract}

{We present the multi-band photometric  and low-resolution spectroscopic analysis of four W Ursae Majoris eclipsing binaries (EWs) - J080510.1+141528  (hereinafter as J0805a), J080516.3+143138  (hereinafter as J0805b), J143358.7+053953  (hereinafter as J1433), and J143458.4+054143  (hereinafter as J1434). The multi-band ground based photometric data are collected using the 1.3-m Devasthal Fast Optical Telescope (DFOT) while we also make use of TESS photometric observations. The spectroscopic analysis is based on the low-resolution observations by 4-m Large Sky Area Multi-Object Fiber Spectroscopic Telescope (LAMOST). The ephemeris of these systems are updated using the photometric data from TESS and other photometric surveys. The system J0805b shows a secular change in the orbital period with a period change rate of 4.2 ($\pm$ 0.1) $\times10^{-7}$ days per year while the orbital period change rate for J1433 is calculated as -1.1 ($\pm$ 0.1) $\times10^{-6}$ days per year. The mass-transfer rate for J0805b is found to be dM$_{1}$/dt = -1.56($\pm$0.07) $\times 10^{-6}$ M$_{\odot}$/year and dM$_{1}$/dt = -7.95($\pm$0.87) $\times 10^{-7}$ M$_{\odot}$/year for J1433. All the systems have inclination > 79$^{\circ}$ except J1433 which has inclination of 72.8$^{\circ}$. The mass-ratios (less massive to more massive component) for these targets are < 0.5. All the system except J0805b are A-subtype contact binaries. The absolute parameters of the systems are determined using GAIA DR3 parallax and reddening information. The LAMOST spectra are analyzed using spectral subtraction technique. A small excess emission is detected for J1433 in H$_{\alpha}$ and H$_{\beta}$ region. The systems are plotted on Hertzsprung–Russell (HR) diagram and compared with previously studied systems. The mass-ratio vs radius-ratio relation is also investigated for these systems.

}
 

\end{abstract}

   
\section{Introduction} \label{sec:intro}
The population of known eclipsing binaries (EBs) has increased tremendously in last few decades due to multiple surveys such as the Optical Gravitational Lensing Experiment (OGLE, \citealt{1992AcA....42..253U}), Robotic Optical Transient Search Experiment (ROTSE, \citealt{2000AJ....119.1901A}), All Sky Automated Survey (ASAS, \citealt{2018MNRAS.477.3145J}), Super Wide Angle Search for Planets (SuperWASP, \citealt{2010A&A...520L..10B}), Catalina Sky Surveys (CSS, \citealt{2014ApJS..213....9D}), and Zwicky Transient Facility (ZTF, \citealt{2019PASP..131a8002B}). This list is further expanded by space based surveys like Kepler \citep{2016RPPh...79c6901B} and Transiting Exoplanet Survey Satellite (TESS, \citealt{2015JATIS...1a4003R}) which have collected high signal to noise ratio (SNR) data with continuous coverage of sky for months or even years. The importance of EBs to determine stellar parameters is well known, but these long term observations are advantageous to detect and study the longer time-scale phenomena, including magnetic activity cycles and members of multiple system \citep{2020A&A...635A..89M, 2022MNRAS.509..246H, 2022ApJS..262...10K}. With the help of available data and advanced data analysis tools, we have markedly improved our understanding of the origin, structure, and evolution of EBs \citep{2017A&A...608A..62H, 2023MNRAS.519.3546S, 2023MNRAS.521..677P, 2024MNRAS.528.5703S}.

A major fraction of the detected EBs from these surveys belongs to the contact binary category. Contact binaries (CBs) consist of two late type stars with a common convective envelope around them due to the overfilled Roche lobes \citep{1956AnAp...19..298K, 1967AJ.....72S.309L}. Most of the CBs have circular and synchronous orbit. The CB light curves (LCs) show continuous flux variation caused by the distorted components and have orbital period ($P_{orb}$) less than 1 day.  Because of the common envelope geometry and high speed rotation of the components, increased magnetic and chromospheric activity are observed in these systems. These effects generate star-spots, flares, asymmetric distribution of circum-binary matter or Coriolis heating in the binary photosphere \citep{2017MNRAS.465.4678M, 2021ApJS..254...10L, 2022ApJS..262...10K} which are observed in form of asymmetry in the CB LCs. This LC asymmetry is known as O'Connell effect \citep{1951PRCO....2...85O}. The O'Connell effect in the present EB models can be explained by the surface brightness non homogeneity in the form of cool/bright spots. With the help of good quality long term observations, it is also possible to determine multiple parameters of spots and track down their movement on the stellar surface with time.
 
The CBs are believed to be formed from detached eclipsing binaries (DEBs) with $P_{orb}$ less than a few days. Multiple mechanisms play important role to bring both the binary components close to each other and transform DEBs into CBs. The important mechanisms which play crucial role in this process are tidal friction, mass transfer between the components, magnetic braking, and Kozai cycles within triple systems. Theoretical studies by \cite{1977MNRAS.179..359R} suggest the end of CBs in the form of a merger event. In fact, V1309 Scorpii a luminous red nova, confirms these theories \citep{2011A&A...528A.114T, 2014ApJ...786...39N} 

In this work, a multi-band photometric analysis of four CBs is presented. Apart from the absolute parameters, we determined the period change and mass transfer rates for these systems. We also analyzed the low-resolution LAMOST spectra of these four CBs, The first system J0805a ($\alpha_{2000}=08^{h}05^{m}10^{s}.1$, $\delta_{2000}=+14^{\circ}15^{\prime}28.2^{\prime\prime}$) was reported and categorized as a CB by the Catalina Surveys Data Release-1 (CSDR1, \citealt{2014ApJS..213....9D}). This system was again mentioned in the ATLAS data release \citep{2018AJ....156..241H}. The second system J0805b ($\alpha_{2000}=08^{h}05^{m}16^{s}.3$, $\delta_{2000}=+14^{\circ}31^{\prime}38.0^{\prime\prime}$) was reported by multiple surveys including Northern Sky Variability Survey (NSVS), CRTS and ATLAS \citep{2004AJ....127.2436W, 2014ApJS..213....9D, 2018AJ....156..241H}. Rest of the systems J1433 ($\alpha_{2000}=14^{h}33^{m}58^{s}.7$, $\delta_{2000}=+05^{\circ}39^{\prime}53.4^{\prime\prime}$) and J1434 ($\alpha_{2000}=14^{h}34^{m}58^{s}.5$, $\delta_{2000}=+05^{\circ}41^{\prime}43.3^{\prime\prime}$)  were reported by ASAS \citep{2002AcA....52..397P} in the beginning but also observed by CRTS, ZTF \citep{2014ApJS..213....9D, 2020ApJS..249...18C}. Table~\ref{tar_info} includes the basic information about these targets.

The paper is divided into multiple sections. The photometric and spectroscopic data related details are given in Section~\ref{Data}. The process of period estimation and period change are discussed in Section~\ref{orpe}. The LC modeling and the parameters determination process are discussed in Section~\ref{Ana} and Section~\ref{phy_para}, respectively. The Section~\ref{ma_tr} discusses the underlying mechanisms for observed period change in these systems. The low-resolution spectra analysis is given in Section~\ref{ch_ac}. In the end, Section~\ref{discu} discusses the final results and conclusion of this work.

\begin{table}[!ht]
\caption{Basic information about the sources taken from different surveys. The period and average V-band magnitude are taken from \cite{2014ApJS..213....9D}, $A_{V}$ is given by \cite{2011ApJ...737..103S}, (J-K) is taken from 2MASS survey \citep{2006AJ....131.1163S} and the parallax is from GAIA \citep{2021A&A...649A...1G}.}   
\centering              
\label{tar_info}
\scriptsize         
\begin{tabular}{p{.25in}p{.35in}p{.37in}p{0.3in}p{0.2in}p{0.3in}p{0.2in}p{0.25in}}
\hline\hline     
Source& RA         &DEC          & Period   & V     & E(B-V) & J-K  & Parallax \\
      & (J2000)    &(J2000)      & (days)   & (mag) & (mag)  & (mag)& (mas)    \\
\hline \hline
J0805a& 08:05:10.1 & +14:15:28.2 & 0.321164 & 15.68 & 0.0302 & 0.53 & 0.6082   \\
J0805b& 08:05:16.3 & +14:31:38.1 & 0.248650 & 13.80 & 0.0259 & 0.55 & 2.0199   \\

J1433 & 14:33:58.7 & +05:39:53.4 & 0.352754 & 13.37 & 0.0263 & 0.41 & 1.3557   \\
J1434 & 14:34:58.5 & +05:41:43.3 & 0.369354 & 12.92 & 0.0299 & 0.33 & 1.4972   \\
\hline\hline               
\end{tabular}
\end{table}

%
\section{Observations}\label{Data}
\subsection{Photometry}\label{Photo}
The multi-band photometric data for all the four sources were collected using the 1.3-m Devasthal Fast Optical Telescope (DFOT) in BVRI-bands. The telescope is located in Devasthal, Nainital and operated by Aryabhatta Research Institute of Observational Sciences (ARIES). The telescope uses a 2048×2048 pixels CCD and high efficiency transmission filters. At a time, the motorized filter wheel can operate with 8 filters out of 13 available filters (broad-band filters UBVRI, SDSS filters ugriz and narrow-band filters H-alpha, O-III, S-II). The operating temperature, gain and the read noise of the CCD was set to -80$^{\circ}$ C, $2e^{-} ADU^{-1}$, and $7.5e^{-}$, respectively, during observations. Other details about the telescope system and available instruments can be found in \cite{2022JAI....1140004J}. The 2048×2048 pixels CCD allows a field of view (FoV) of $\sim 18^{\arcmin}\times18^{\arcmin}$, which is good enough to provide a sufficient number of comparison stars required for the differential photometry. The raw data files are cleaned with PyRAF following the standard steps like bias subtraction, flat-fielding, and cosmic ray removal. To estimate the flux and generate the final differential LCs, a general image processing package AstroImageJ (AIJ) was used \citep{2017AJ....153...77C}. The complete photometric observations for the sources from DFOT are provided in the Table~\ref{log_phot}.

The targets were also observed by Transiting Exoplanet Survey Satellite (TESS) mission during different sector visits. TESS was launched in 2018 with an aim to find Earth-sized planets around bright stars. After finishing its 2 year prime mission in 2020, TESS has completed first extended mission and still in orbit, scanning the Sector 90 as a part of next extended mission. The data from TESS mission are available for use after being processed by the Science Processing Operations Center (SPOC) and the Quick-Look Pipeline (QLP). TESS SPOC and TESS High Level Science Products-QLP (HLSP-QLP) are available at the Mikulski Archive for Space Telescopes (MAST) portal \footnote{https://mast.stsci.edu}. We downloaded the LCs for J0805b, J1433, and J1434 from MAST portal while data was unavailable for J0805a. In case of J0805a, the LCs from different TESS sector observations were generated using a tool named TESS-Gaia Light Curve (TGLC) \footnote{https://pypi.org/project/tglc/0.5.1/}. TGLC produces accurate LCs for stars by utilizing astrometric and photometric information from Gaia mission \citep{2023AJ....165...71H}. TGLC can generate LCs for stars with TESS magnitudes less than 16 using effective point spread function (ePSF). The irregularly contamination in TESS frames may be present after the end of each $\sim$14 day observation span due to the scattered light from Earth and Moon. The frames with this kind of high unexpected background are flagged with "low-quality" label in TGLC output time-series. The details about TESS observations are given in the Table~\ref{log_photb}.

\begin{table*}
\caption{DFOT observations for the objects, including Julian Date (JD), relative flux, and relative flux error for each photometric band (B, V, R, I). Observations for each object are provided in four separate blocks from top (for J0805a) to bottom (for J1434). This table provides a shortened version, with the full dataset available in electronic form.}
\centering
\label{log_phot}
\scriptsize

\begin{tabular}{l l l l l l l l l l l l}   
\hline\hline
    JD  & Relative   & Error      &    JD  & Relative   & Error      &    JD  & Relative   & Error      &    JD  & Relative   & Error      \\
        & B-Flux     & B-Flux     &        & V-Flux     & V-Flux     &        & R-Flux     & R-Flux     &        & I-Flux     & I-Flux     \\
\hline\hline           
2459949.196729 & 0.168 & 0.006 & 2459949.200155 & 0.162 & 0.003 & 2459949.202875 & 0.183 & 0.003 & 2459949.204901 & 0.194 & 0.004    \\
2459949.198175 & 0.162 & 0.006 & 2459949.201601 & 0.173 & 0.004 & 2459949.203626 & 0.184 & 0.003 & 2459949.205652 & 0.187 & 0.004    \\
2459949.208188 & 0.152 & 0.006 & 2459949.210908 & 0.165 & 0.005 & 2459949.212991 & 0.172 & 0.003 & 2459949.215306 & 0.181 & 0.004    \\
2459949.209633 & 0.163 & 0.006 & 2459949.211659 & 0.165 & 0.005 & 2459949.213742 & 0.174 & 0.003 & 2459949.216057 & 0.198 & 0.004    \\
2459949.218663 & 0.151 & 0.006 & 2459949.221626 & 0.166 & 0.004 & 2459949.223628 & 0.187 & 0.003 & 2459949.225677 & 0.191 & 0.004    \\
-- & -- & -- & -- & -- & -- & -- & -- & -- & -- & -- & --    \\
\hline
2459949.196729 & 0.808 & 0.008 & 2459949.200155 & 0.921 & 0.005 & 2459949.202875 & 1.068 & 0.005 & 2459949.204901 & 1.211 & 0.007    \\
2459949.198175 & 0.827 & 0.008 & 2459949.201601 & 0.957 & 0.005 & 2459949.203626 & 1.086 & 0.005 & 2459949.205652 & 1.227 & 0.007    \\
2459949.208188 & 0.981 & 0.009 & 2459949.210908 & 1.098 & 0.007 & 2459949.212991 & 1.206 & 0.005 & 2459949.215306 & 1.337 & 0.007    \\
2459949.209633 & 0.992 & 0.009 & 2459949.211659 & 1.091 & 0.007 & 2459949.213742 & 1.208 & 0.005 & 2459949.216057 & 1.336 & 0.007    \\
2459949.218663 & 1.069 & 0.009 & 2459949.221626 & 1.171 & 0.007 & 2459949.223628 & 1.277 & 0.006 & 2459949.225677 & 1.427 & 0.007    \\
-- & -- & -- & -- & -- & -- & -- & -- & -- & -- & -- & --    \\
\hline
2459581.438779 & 0.492 & 0.004 & 2459274.342650 & 0.534 & 0.004 & 2459274.343958 & 0.561 & 0.004 & 2459274.345382 & 0.574 & 0.003    \\
2459581.439183 & 0.492 & 0.004 & 2459274.342939 & 0.542 & 0.004 & 2459274.344247 & 0.561 & 0.003 & 2459274.345670 & 0.577 & 0.003    \\
2459581.439587 & 0.494 & 0.004 & 2459274.343227 & 0.535 & 0.004 & 2459274.344535 & 0.559 & 0.003 & 2459274.345958 & 0.572 & 0.003    \\
2459581.439991 & 0.486 & 0.004 & 2459274.347164 & 0.545 & 0.004 & 2459274.348530 & 0.567 & 0.003 & 2459274.349965 & 0.579 & 0.003    \\
2459581.440395 & 0.491 & 0.004 & 2459274.347453 & 0.538 & 0.004 & 2459274.348818 & 0.559 & 0.003 & 2459274.350253 & 0.581 & 0.003    \\
-- & -- & -- & -- & -- & -- & -- & -- & -- & -- & -- & --    \\
\hline
2459581.438779 & 0.799 & 0.005 & 2459274.342650 & 0.786 & 0.004 & 2459274.343958 & 1.077 & 0.005 & 2459274.345382 & 0.772 & 0.004    \\
2459581.439183 & 0.793 & 0.005 & 2459274.342939 & 0.789 & 0.004 & 2459274.344247 & 0.793 & 0.003 & 2459274.345670 & 0.779 & 0.004    \\
2459581.439587 & 0.806 & 0.005 & 2459274.343227 & 0.787 & 0.004 & 2459274.344535 & 0.790 & 0.003 & 2459274.345958 & 0.776 & 0.004    \\
2459581.439991 & 0.813 & 0.005 & 2459274.347164 & 0.781 & 0.004 & 2459274.348530 & 0.783 & 0.003 & 2459274.349965 & 0.767 & 0.004    \\
2459581.440395 & 0.811 & 0.005 & 2459274.347453 & 0.784 & 0.004 & 2459274.348818 & 0.778 & 0.003 & 2459274.350253 & 0.764 & 0.004    \\
-- & -- & -- & -- & -- & -- & -- & -- & -- & -- & -- & --    \\
\hline\hline 
 \end{tabular}       
 \end{table*}

%
\begin{table}
\caption{The TESS observation log for our targets.}
\centering
\label{log_photb}
\scriptsize

\begin{tabular}{p{.35in}p{.2in}p{.55in}p{0.55in}p{0.35in}p{0.35in}}
\hline
\hline
Object& Sector &  Start BJD  &  End BJD    & No. of   & Exposure  \\
      &        & (2400000+)  & (2400000+)  & Frames   & time (sec)\\
\hline
\hline
      &   44   & 59500.1901  & 59524.4355  &   3278   & 600       \\
      &   45   & 59525.5117  & 59550.6251  &   3450   & 600       \\
J0805a&   46   & 59551.5693  & 59578.7031  &   3707   & 600       \\
      &   71   & 60234.0301  & 60259.9679  &  10932   & 200       \\
      &   72   & 60260.1804  & 60285.5854  &  10701   & 200       \\
\hline
      &   44   & 59500.1901  & 59524.4355  &   3278   & 600       \\
      &   45   & 59525.5117  & 59550.6251  &   3450   & 600       \\
J0805b&   46   & 59551.5693  & 59578.7031  &   3707   & 600       \\
      &   71   & 60234.0301  & 60259.9679  &  10932   & 200       \\
      &   72   & 60260.1804  & 60285.5854  &  10701   & 200       \\
\hline
J1433 &   51   & 59692.9532  & 59717.5363  &   3399   & 600       \\
\hline
J1434 &   51   & 59692.9532  & 59717.5363  &   3399   & 600       \\
\hline\hline 
 \end{tabular}       
 \end{table}

%
\begin{figure*}[!ht]
\begin{center}
\subfigure{\includegraphics[width=15cm,height=6cm]{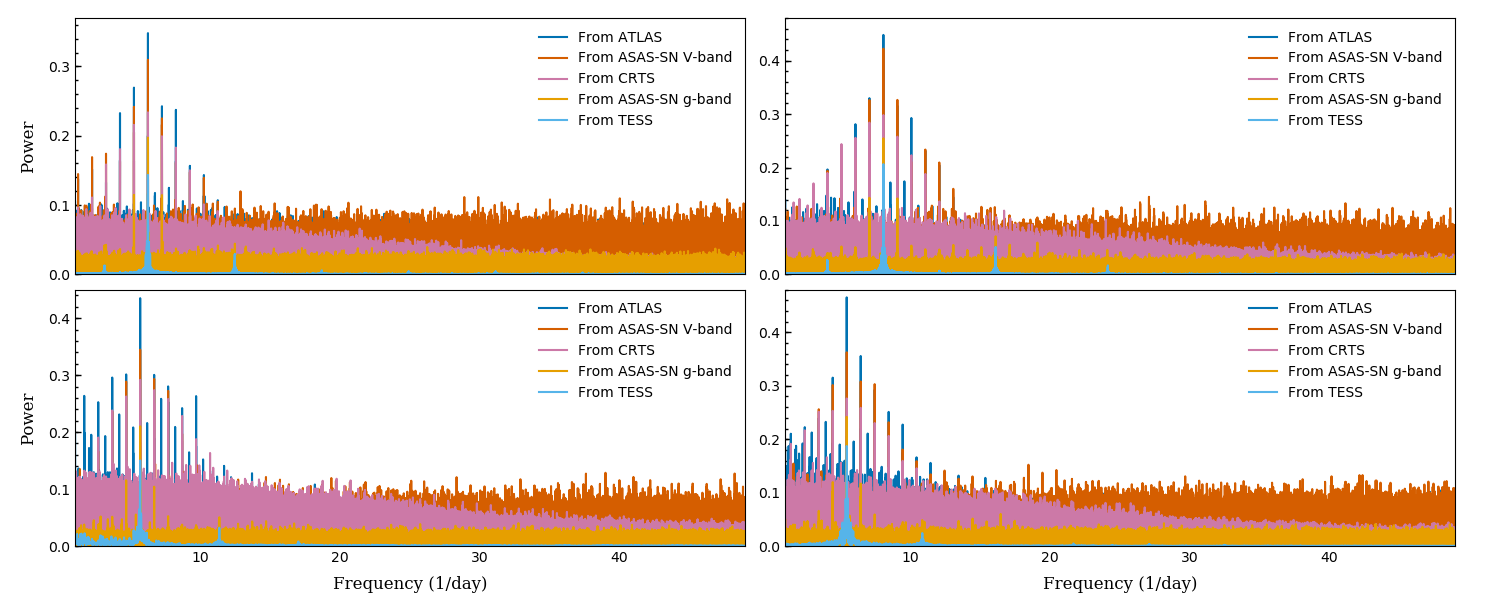}}
\caption{The power spectra for all the sources, with each dataset represented by a different color.}
\label{periodo}
\end{center}
\end{figure*}
\subsection{Spectroscopy}\label{Spec}
Large Sky Area Multi-Object Fiber Spectroscopic Telescope (LAMOST) or the GuoShouJing Telescope, is a special quasi-meridian reflecting Schmidt telescope with a wide FoV of 5$^{\circ}$. It is a multi-fiber Spectroscopic Telescope which uses 16 spectrographs and there are 250 fibers connected to each spectrograph. With an exposure time of 1.5 hours, it can reach up to the limiting magnitude of 20.5 with resolution R=500 \citep{2015RAA....15.1095L}. In February 2024, LAMOST published the 12th data release. LAMOST data from first 10 data releases (DR) are publicly available. Till DR10, LAMOST has collected $\sim$ 11.8 million low resolution spectra (LRS) and $\sim$ 10.5 million medium resolution spectra (MRS). LAMOST LRS database consists of $\sim$ 11 million star spectra, $\sim$ 261 thousand galaxy, and $\sim$ 79 thousand QSO spectra. LAMOST also provides multiple parameters such as temperature ($T_{eff}$), surface gravity (log g), metallicity (Fe/H), radial velocity (RV), etc. calculated with the help of the LAMOST stellar parameter pipeline (LASP; \citealt{2014IAUS..306..340W}). Our targets are observed by LAMOST in the low resolution mode \footnote{https://www.lamost.org/dr9/}.  Some important information about the targets and  LASP parameters are given in the Table~\ref{tar_lamost}. 

\begin{table}[!ht]
\caption{Parameters of targets from the LAMOST data}             
\label{tar_lamost}
\centering   
\scriptsize       
\begin{tabular}{p{.39in}p{.49in}p{.2in}p{0.28in}p{0.25in}p{0.27in}p{0.25in}}
\hline\hline 
Targets & Date        &$T_{eff} $& Sub & logg  & Fe/H   & SNR \\
        &             &    (K)   &class&       & (dex)  &     \\
\hline\hline
J0805a  & 21-12-2015  & 5608     & G3  & 3.957 & -0.170 & 55  \\ 
	
\hline
J0805b  & 21-03-2014  & 5249     & G7  & 4.084 & -0.641 & 121 \\ 
        & 21-12-2015  & 5226     & G8  & 4.113 & -0.646 & 244 \\ 

\hline
J1433   & 27-01-2016  & 5687     & G3  & 3.924 & -0.043 & 58  \\ 
        
\hline
J1434   & 17-03-2013  & 6014     & F9  & 3.944 &  0.041 & 77  \\ 

\hline\hline                
\end{tabular}
\end{table}

%
\begin{figure*}[!ht]
\begin{center}
\subfigure{\includegraphics[width=17cm,height=6cm]{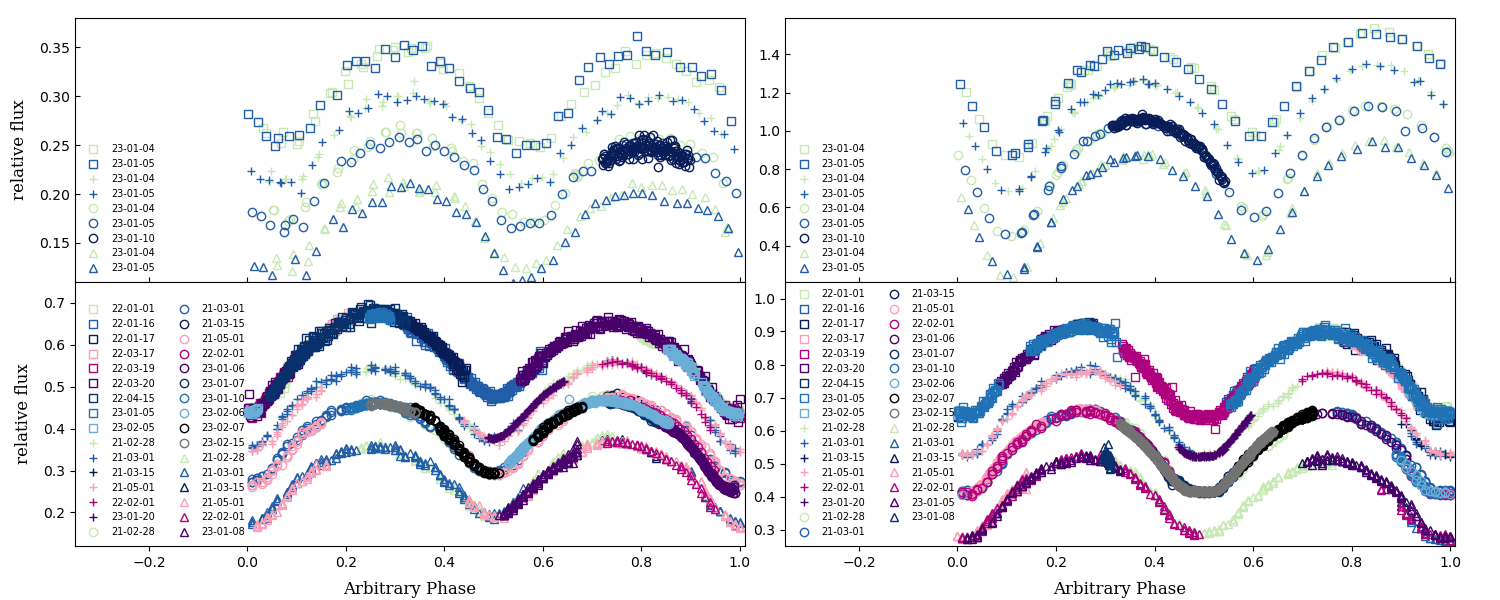}}
\caption{Phase folded LC for all the targets are shown. Different colors are used for different dates of observations and different symbols are used for different photometric bands.}
\label{lc_obs}
\end{center}
\end{figure*}
%
\section{Period and Period Variation}\label{orpe}
As mentioned in the Section~\ref{sec:intro}, these objects were also observed in previous surveys. We collected the available photometric time-series for these objects from different surveys like CRTS, ASAS-SN, ATLAS, TESS, ZTF, GAIA, AAVSO Photometric All-Sky Survey (APASS), and KWS (Kamogata/Kiso/Kyoto Wide-field Survey). Given the longer observational coverage and higher data quality, we used only CRTS, ASAS-SN, ATLAS, and TESS photometric observations for period analysis. For the initial period determination, popular time series analysis tool PERIOD04 was used. It is a Java and C++ based program, which helps to analyze long time-series data with the help of discrete Fourier transform. It provides a simple Graphical User Interface (GUI) to use. User can fit multiple frequencies and estimate the uncertainties with the help of Markov chain Monte Carlo (MCMC) analysis \citep{2004IAUS..224..786L, 2005CoAst.146...53L}. Figure~\ref{periodo} shows the power spectra for all the sources based on different datasets. The peak frequencies obtained from the periodogram analysis are given in Table~\ref{peak_info}. We calculated the average of all of the frequencies obtained from different datasets for each source and determined the $P_{orb}$ for each system. The $P_{orb}$ for J0805a, J0805b, J1433, and J1434 are found to be 0.3211636, 0.2486499, 0.3527459, and 0.3693600 days. The multi-band LCs from DFOT are phase folded using calculated $P_{orb}$. The LCs are shown in Figure~\ref{lc_obs}. The B, V, R, and I photometric bands are represented by open squares, plus symbol, open circles, and open triangles.

\begin{table}[!ht]
\caption{Times of minima (ToMs) brightness for J0805a, J0805b, J1433 and J1434 as calculated using different data sources}
\label{OC_info}
\centering
\scriptsize
\begin{tabular}{l c c c c c}   
\hline\hline
ID     & $BJD_{o}$   & Error &  Cycle  & $(O-C)_{1}$& Source \\
       & (2400000+)  &       &         &  (days)    &        \\
\hline\hline
       &             &       &         &            &        \\
J0805a & 2459500.538 & 0.001 & -1397.0 & -0.003     & T      \\
J0805a & 2459500.861 & 0.001 & -1396.0 & -0.001     & T      \\
   -   &       -     &   -   &    -    &     -      &        \\
J0805b & 2458508.755 & 0.001 & -5794.0 & +0.003     & AC     \\
J0805b & 2459249.729 & 0.001 & -2814.0 & +0.002     & AC     \\
   -   &       -     &   -   &    -    &     -      &        \\
J1433  & 2459633.370 & 0.002 & -0185.0 & +0.008     & D      \\
J1433  & 2459336.350 & 0.002 & -1027.0 & -0.0002    & D      \\
   -   &       -     &   -   &    -    &     -      &        \\
J1434  & 2459274.7857& 0.0002& -1133.0 & +0.0005    & D      \\
J1434  & 2459589.8463& 0.0001& -0280.0 & -0.0021    & D      \\
   -   &       -     &   -   &    -    &     -      &        \\
\hline\hline             
\end{tabular}
\end{table}

\subsection{J0805a}\label{J0805a_per_stu}

\begin{figure*}[!ht]
\label{oc_J0805}
\begin{center}
\subfigure{\includegraphics[width=7cm,height=4cm]{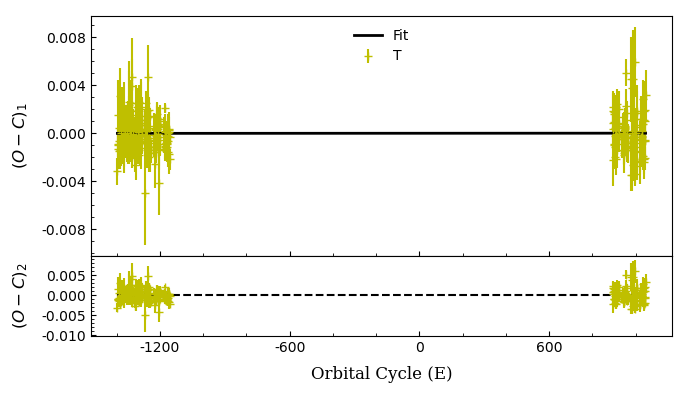}}
\subfigure{\includegraphics[width=7cm,height=4cm]{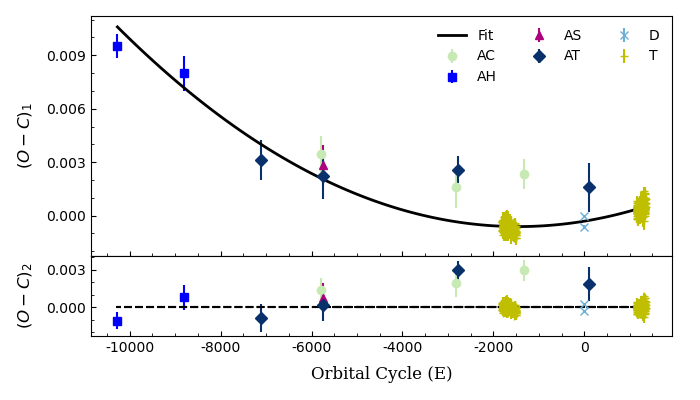}}
\caption{O-C diagrams for J0805a and J0805b with linear and quadratic fit, respectively. Primary ToMs obtained from different datasets are represented by different symbols and colors. The residuals of the fits are displayed in the lower panel of each plot.}
\end{center}
\end{figure*}
J0805a was observed by TESS mission during sector 44, 45, 46, 71, and 72 visits. The time series data are available with an exposure time of 600 seconds for first three visits and 200 seconds for last two visits. With the help of parabola fitting we estimated the ToMs for TESS observations. Using these multiple sector data, a total of 291 Primary ToMs were determined (s44: 54 ToMs, s45: 67 ToMs, s46: 56 ToMs, s71: 59 ToMs, s72: 55 ToMs). The linear ephemeris was updated for the system and determined as follows:

\begin{equation}
\label{li_J0805a}
\begin{aligned}
BJD_{\circ} &=2459949.2088(\pm 0.0001) \\
&+0.32116501(\pm 0.00000008)\times E
\end{aligned}
\end{equation} 
Here, $BJD_{\circ}$ is the TOM for E$^{th}$ orbital cycle and E is an integer representing orbital cycle number. Using Equation~\ref{li_J0805a}, we generated the O-C diagram for the system. The O-C diagram and the fitted straight line is shown in Figure~\ref{oc_J0805} (left panel). The O-C diagram follows a linear trend which can be represented by:
\begin{equation}
\label{li_J0805a_oc}
\begin{aligned}
(O-C)_{1} &=(-0.04\pm9.65) \times10^{-5} \\
&+(0.66\pm8.27) \times10^{-8}  \times E
\end{aligned}
\end{equation}
 
Statistically, this fit is equivalent to (O-C)=0, as both the slope and intercept are close to zero. This suggests that the $P_{orb}$ of the system J0805a is stable and does not exhibit any significant changes over the observed data range.

\begin{table}[!ht]
\caption{List of peak frequencies (1/day) determined by PERIOD04 periodogram for each source using different datasets.}
\label{peak_info}
\centering
\scriptsize
\begin{tabular}{l c c c c}   
\hline\hline
Survey  & J0805a     & J0805b     & J1433      & J1434      \\
\hline\hline
ATLAS   & 6.22737473 & 8.04342974 & 5.66979723 & 5.41478003 \\
ASAS-g  & 6.22737413 & 8.04344579 & 5.66980107 & 5.41477379 \\
ASAS-V  & 6.22738895 & 8.04345495 & 5.66973890 & 5.41479531 \\
CRTS    & 6.22732359 & 8.04342745 & 5.66969141 & 5.41481672 \\
TESS    & 6.22731938 & 8.04342321 & 5.66999207 & 5.41468991 \\
\hline\hline       
\end{tabular}
\end{table}

\subsection{J0805b}\label{J0805b_per_stu}
Like J0805a, J0805b was also observed by TESS in 5 sectors including 44, 45, 46, 71, and 72. Following the similar parabola fitting technique, we derived 319 primary ToMs (s44: 72 ToMs, s45: 45 ToMs, s46: 43 ToMs, s71: 74 ToMs, s72: 85 ToMs) for J0805b using TESS observations. The system was also observed by ASAS-SN survey telescopes in V and g-bands. The ASAS-SN LCs for the source were generated and downloaded from the ASAS-SN Sky Patrol website \citep{2014ApJ...788...48S, 2023arXiv230403791H}. The system was observed in V-band from February, 2012 to November, 2018 and from October, 2017 to March, 2024 in g-band. The ASAS-SN telescopes are located in Hawaii, Texas, South Africa, and Chile. We filtered the data on the basis of ASAS-SN observatories. As ASAS-SN telescopes monitor the entire sky with a cadence of $\leq$ 1 day, their data are not enough to determine the TOMs using parabola fitting. Instead of using each orbital cycle LC for ASAS-SN data, we divided the each observatory data into $\sim$300 days data subsets. The TOMs for each of data subsets were determined using PHOEBE model fitting. On the basis of determined ToMs, the new linear ephemeris was determined as follows:

\begin{equation}
\label{li_J0805b}
\begin{aligned}
BJD_{\circ} &=2459949.42897(\pm0.00006) \\
&+0.24865002(\pm0.00000003)\times E
\end{aligned}
\end{equation}

The updated quadratic ephemeris for J0805b can be represented by:
\begin{equation}
\label{qu_J0805b}
\begin{aligned}
BJD_{\circ} &=2459949.42866(\pm0.00002)\\
&+0.24865042(\pm0.00000002) \times E \\
&+(1.44\pm0.04) \times 10^{-10} \times E^{2}
\end{aligned}
\end{equation}

The O-C diagram generated using this updated linear ephemeris is shown in the right panel of Figure~\ref{oc_J0805}. The different ASAS-SN observatories are shown with different color symbols in the plot. The labels AC, AH, AS, AT, D, and T represent ASAS-SN Chile, ASAS-SN Hawaii, ASAS-SN SAAO, ASAS-SN Texas, DFOT, and TESS, respectively. The non-linear variation of the O-C diagram can be represented by following equation: 
\begin{equation}
\label{qu_J0805b_oc}
\begin{aligned}
(O-C)_{1} &=-0.00031(\pm0.00002)\\
&+(4.2\pm0.2)\times 10^{-7} \times E \\
&+(1.44\pm0.04) \times 10^{-10} \times E^{2}
\end{aligned}
\end{equation}
The non-linear trend observed in the O-C diagram of the system J0805b is expected due to the changing $P_{orb}$ of the system. With the help of O-C diagram, we were able to calculate the $P_{orb}$ change rate for the system as 4.2 ($\pm$ 0.1) $\times10^{-7}$ days per year.
\subsection{J1433}\label{J1433a_per_stu}
The system J1433 was observed by TESS during sector 51 visit. It was observed continuously for $\sim$ 24.5 days with an exposure time of 600 seconds by TESS mission. Using parabola fitting, a total of 35 ToMs were determined from TESS data. Similar to J0805b, we collected ASAS-SN observations for J1433 from the ASAS-SN Sky Patrol website. The ASAS-SN observations for J1433 were available from February, 2012 to August, 2018 in V-band and September, 2017 to March, 2024 in g-band. The system was also observed by ZTF survey. The ZTF observations for J1433 from March, 2018 to August, 2023 were also collected. Using the PHOEBE modeling, we determined 13 primary ToMs from ASAS-SN data and 6 primary ToMs from ZTF data. Apart from 13 ToMs from ASAS-SN and 6 ToMs from ZTF, we also determined 4 primary ToMs from DFOT observations. The updated ephemeris on the basis of these observations was determined as follows:
\begin{equation}
\label{li_J1433a}
\begin{aligned}
BJD_{\circ} &=2459698.6203(\pm0.0003) \\
&+0.3527462(\pm0.0000002)\times E
\end{aligned}
\end{equation} 

The updated quadratic ephemeris for J1433 can be represented by:
\begin{equation}
\label{qu_J1433}
\begin{aligned}
BJD_{\circ} &=2459698.6201(\pm0.0002)\\
&+0.3527434(\pm0.0000004) \times E \\
&-(5.4\pm0.6) \times 10^{-10} \times E^{2}
\end{aligned}
\end{equation}

\begin{figure*}[!ht]
\begin{center}
\label{oc_J1433}
\subfigure{\includegraphics[width=7cm, height=4cm]{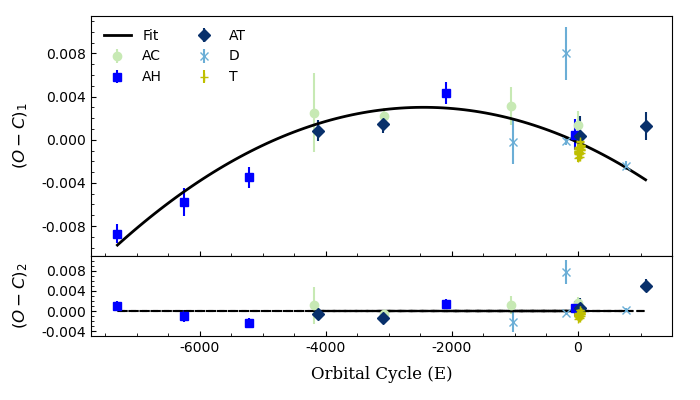}}
\subfigure{\includegraphics[width=7cm, height=4cm]{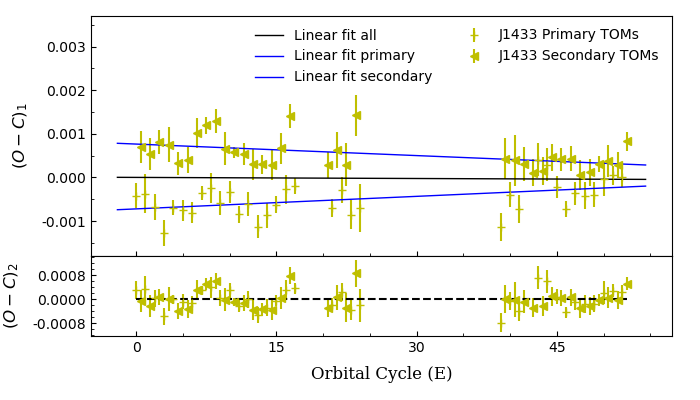}}
\caption{The O-C diagram for J1433 with quadratic is shown in the left panel, the right panel shows the TESS part of O-C diagram with primary and secondary ToMs. The different types of symbols and colors in the plots represent different data sources.}
\end{center}
\end{figure*}

The left panel of Figure~\ref{oc_J1433} shows the O-C diagram for J1433 with a quadratic fit. The following equation expresses the best quadratic fit to the O-C data: 
\begin{equation}
\label{qu_J1433a_oc}
\begin{aligned}
(O-C)_{1} &=-0.0002(\pm0.0002)\\
&-(2.6\pm0.4)\times 10^{-6} \times E \\
&-(5.4\pm0.7) \times 10^{-10} \times E^{2}
\end{aligned}
\end{equation}
The non-linear O-C variation in the system J1433 can be associated with the changing $P_{orb}$ with time. The rate of period change for J1433 is determined as -1.1 ($\pm$ 0.1) $\times10^{-6}$ days per year. The right panel of Figure~\ref{oc_J1433} shows the TESS portion of the complete O-C diagram. It reveals short-term variations, along with distinct linear trends (following blue lines in the plot) for the primary and secondary O-C values. As the system was observed by TESS mission only one time, determining the exact cause of the variation in the O-C diagram is challenging; however, it may be attributed to the presence of spots on the components of the system \citep{2013ApJ...774...81T, 2015MNRAS.448..429B}.
%

\subsection{J1434}\label{J1434b_per_stu}
The system J1434 is $\sim$ $15^{'}$ away from J1433, so, it was also observed by TESS during same sector 51 coverage. The system was observed by ASAS-SN survey from February, 2012 to August, 2018 in V-band and September, 2017 to March, 2024 in g-band. We also used ZTF data in the zr and zg bands, collected between March 2018 and August 2023. In total we determined 59 primary ToMs for the system. We calculated 37 ToMs from TESS , 11 from ASAS-SN, 8 from ZTF and 3 from DFOT observations. The updated linear ephemeris for the system on the basis of TESS observations is calculated as follows:

\begin{equation}
\label{li_J1434b}
\begin{aligned}
BJD_{\circ} &=2459693.2689(\pm0.0002) \\
&+0.3693593(\pm0.0000002)\times E
\end{aligned}
\end{equation} 

\begin{figure*}[!ht]
\begin{center}
\label{oc_J1434}
\subfigure{\includegraphics[width=7cm, height=4cm]{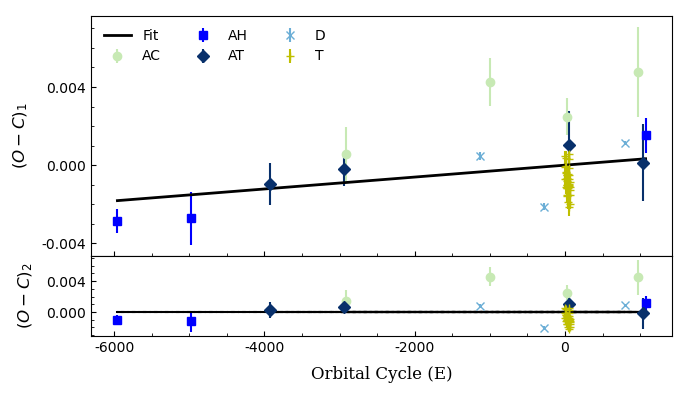}}
\subfigure{\includegraphics[width=7cm, height=4cm]{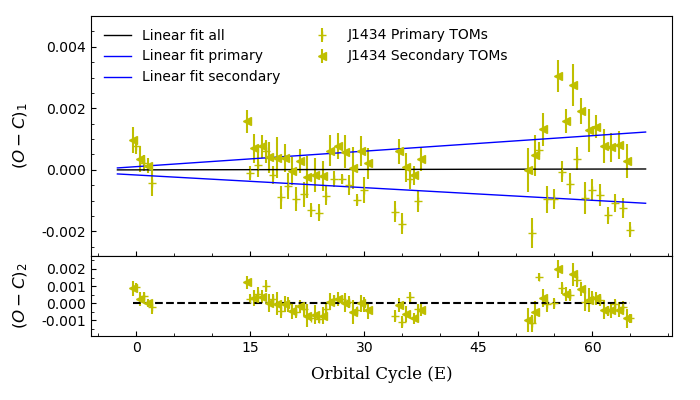}}
\caption{The left panel shows the O-C diagrams for J1434 with a straight line fit while the right panel shows the variation among the TESS part of O-C diagram. The different types of symbols and colors in the plots represent different data sources.}
\end{center}
\end{figure*}

The O-C diagram for J1434 with a linear fit is shown in the right panel of Figure~\ref{oc_J1434}. Due to poor quality of ZTF and ASAS-SN data as compared to TESS, the error bars are large. The following linear equation expresses the best fit: 
\begin{equation}
\label{li_J1434b_oc}
\begin{aligned}
(O-C)_{1} &=(0.003\pm 2.413)\times 10^{-4} \\
& +(3.05\pm1.72) \times 10^{-7} \times E
\end{aligned}
\end{equation}
The above equation~\ref{li_J1434b_oc} indicates the non-variable nature of the $P_{orb}$ of system J1434. Similar to J1433, short-term variations in the O-C diagram of J1434 as observed in TESS data are shown in the right panel of the Figure~\ref{oc_J1434}. Their presence can be linked to the presence or migration of spots on the stellar surface. However, future observations of these systems could provide more definitive insights in this direction.

%
\section{Light Curve Modeling}\label{Ana}
There are several tools available for the analysis of eclipsing binaries (EBs), such as Wilson-Devinney code \citep{1971ApJ...166..605W}, ELLC \citep{2016A&A...591A.111M}, jktebop \citep{2004MNRAS.351.1277S, 2007A&A...467.1215S}, and PHOEBE (PHysics Of Eclipsing BinariEs: \citealt{2005ApJ...628..426P}). In this study, PHOEBE, one of the most widely used software packages, was employed. PHOEBE is built on the well-known WD code but offers enhanced capabilities. Its graphical user interface (GUI) allows users to visualize synthetic data, residuals from fit, track spots on stellar surface, while the PHOEBE-scripter is particularly useful for performing statistical analysis on large datasets and multi-band observations simultaneously. The software includes a range of physical effects, such as simple reflection, limb darkening, gravity darkening, and the incorporation of spots. It also provides access to over 30 pass-bands from different surveys and 8 models for different types of EBs. As these 4 systems were identified as contact binaries in earlier surveys, the 'Over-contact binary not in thermal contact' model was selected for analysis. The following sequence is used for the analysis of EB photometric data:

\begin{table}[!ht]
\caption{The  photometric observations collected for these sources from different catalogs. Both the flux and flux error are measured in units of erg/s/cm$^{2}$/\AA.}             
\label{sed_phot}
\scriptsize
\centering          
\begin{tabular}{p{.25in}p{.4in}p{.19in}p{.19in}p{.19in}p{.19in}p{.19in}p{.19in}p{.19in}p{.19in}}    
\hline\hline     
$\lambda$  & Passband   &  Flux   & Error   &  Flux   & Error   &  Flux   & Error    &  Flux   & Error      \\
(\AA)      &           &x$10^{-14}$&x$10^{-18}$&x$10^{-14}$&x$10^{-18}$&x$10^{-14}$&x$10^{-18}$&x$10^{-14}$&x$10^{-18}$      \\
\hline
3522.16    &  SDSS.U    &  0.098  &  0.081  &  0.322  &  0.178  &  -----  &  -----   &  0.955  &  0.439    \\
4303.59    &  APASS.B   &  0.142  &  2.583  &  0.681  &  11.09  &  1.473  &  14.92   &  2.169  &  31.36   \\
4694.35    &  SDSS.G    &  0.180  &  0.067  &  0.619  &  0.114  &  -----  &  -----   &  -----  &  -----    \\
4702.50    &  APASS.G   &  -----  &  -----  &  -----  &  -----  &  1.621  &  17.47   &  2.367  &  34.88  \\
5437.31    &  APASS.V   &  0.151  &  2.170  &  0.775  &  20.56  &  1.602  &  15.35   &  -----  &  -----    \\
5437.48    &  GAIA3E.BP &  0.141  &  0.340  &  0.849  &  2.127  &  1.314  &  3.649   &  2.203  &  30.03   \\
6175.58    &  APASS.R   &  -----  &  -----  &  -----  &  -----  &  1.438  &  11.79   &  1.970  &  6.066   \\
6177.92    &  SDSS.R    &  0.161  &  0.059  &  0.663  &  0.183  &  -----  &  -----   &  -----  &  -----    \\
6710.93    &  GAIA3E.G  &  0.118  &  0.112  &  0.747  &  0.758  &  1.112  &  0.987   &  1.817  &  0.335    \\
7489.98    &  APASS.I   &  -----  &  -----  &  -----  &  -----  &  1.076  &  12.49   &  1.634  &  1.588   \\
7505.34    &  SDSS.I    &  0.128  &  0.047  &  -----  &  -----  &  -----  &  -----   &  1.315  &  0.121    \\
7979.59    &  GAIA3E.RP &  0.102  &  0.237  &  0.659  &  1.551  &  0.912  &  2.378   &  1.287  &  3.671   \\
8936.17    &  SDSS.Z    &  0.096  &  0.053  &  -----  &  -----  &  -----  &  -----   &  -----  &  -----    \\
12412.1    &  2MASS.J   &  0.046  &  0.126  &  0.318  &  0.674  &  0.424  &  0.938   &  0.604  &  1.335   \\
16497.1    &  2MASS.H   &  0.025  &  0.079  &  0.180  &  0.347  &  0.211  &  0.486   &  0.285  &  0.578    \\
21909.2    &  2MASS.KS  &  0.010  &  0.047  &  0.074  &  0.137  &  0.087  &  0.184   &  0.114  &  0.221    \\
33791.9    &  WISE.W1   &  0.002  &  0.005  &  0.013  &  0.029  &  0.016  &  0.034   &  0.020  &  0.042    \\
46293.0    &  WISE.W2   &  0.001  &  0.002  &  0.004  &  0.008  &  0.005  &  0.009   &  0.006  &  0.011    \\
\hline\hline                 
\end{tabular}
\end{table}

\subsection{Effective Temperature Determination}\label{teff}

\begin{figure}[!ht]
\includegraphics[width=\columnwidth]{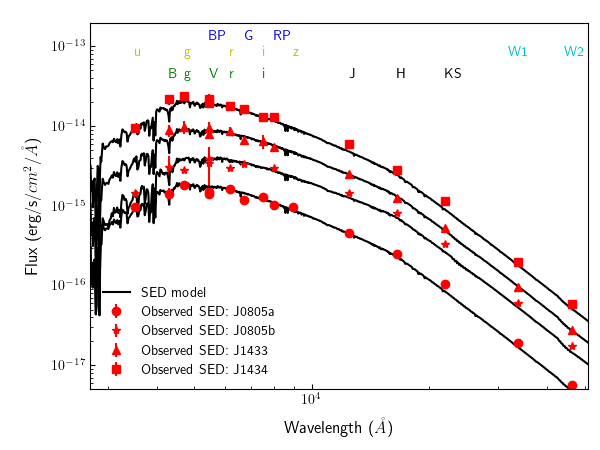}
\caption{The observed SEDs along with SPEEDYFIT synthetic SEDs are shown for all of the sources. For clear visualization, arbitrary vertical shifts are used for each source. The used 2MASS, APASS, GAIA, SDSS, and WISE pass-bands are written around their central wavelength in black, green, blue, yellow, and cyan colored text, respectively.}
\label{all_sed}
\end{figure}

While generating the synthetic fit, PHOEBE needs effective temperature ($T_{eff}$) estimate for one of the components as it scales the $T_{eff}$ of another with respect to the provided $T_{eff}$. Similarly, other parameters like radius or luminosity of components are also temperature dependent. GAIA-DR3 and LAMOST catalog provide the effective temperature ($T_{eff}$) of these targets. We also determined the $T_{eff}$ for each of the systems through SED fitting. A python based package Speedyfit\footnote{https://speedyfit.readthedocs.io/en/latest/} was used for the analysis of the photometric spectral energy distribution of stars \citep{2017A&A...605A.109V, 2018MNRAS.473..693V, 2025ascl.soft02002V}. This can be utilized to access multiple online databases for photometric observations like GALEX, GAIA, Skymapper, APASS, SDSS, Stroemgren-Crawford, 2MASS, and WISE. Currently, Speedyfit includes 4 grids of model atmospheres: Kurucz, Munari, TMAP, and Black Body. Users can add new catalogs as well as other new model atmosphere grids to Speedyfit. Speedyfit uses the Markov chain Monte Carlo technique to fit parameters using the provided information about priors and parameter constraints. Speedyfit can handle both single and binary systems. We used single SED fit for these sources, as both the components in contact binaries have nearly equal $T_{eff}$ with maximum difference in $T_{eff}$ being few hundred kelvin. The reddening and the distance information are taken from \cite{2011ApJ...737..103S} and GAIA DR3, respectively. Figure~\ref{all_sed} shows the observed and synthetic SEDs for these targets. The observed photometry for J0805a, J0805b, J1433, and, J1434 are represented with filled circles, stars, triangles and squares in the plot. The used pass-bands are  indicated at the top of the plot. 2MASS, APASS, GAIA, SDSS, and WISE are shown in black, green, blue, yellow, and cyan colored text, respectively. The photometric data used to generate the SEDs are given in the Table~\ref{sed_phot}.

Other than SED fitting, we collected $T_{eff}$ estimates for these systems from LAMOST and GAIA DR3 surveys. The LAMOST and GAIA DR3 $T_{eff}$ estimates are based on low-resolution spectra (R=1800) and BP/RP spectra, respectively. Using the J and H- band photometry from 2MASS survey, the $T_{eff}$ was determined for each of the systems. The relation is as follows:

\begin{equation}
\label{temp_rela}
\begin{aligned}
T_{eff} =-4369.5(J-H)+ 7188.2
\end{aligned}
\end{equation}

The relation~\ref{temp_rela} is given by \cite{2007MNRAS.380.1230C} using the J-H index information for a sample of $\sim$ 65000 stars from 2MASS data. As reported by \cite{2007MNRAS.380.1230C}, this relation is valid over the $T_{eff}$ range of 4000 to 7000 K. The estimated temperatures using different techniques/surveys are given in the Table~\ref{all_temp}. The average of these multiple $T_{eff}$ estimates is used as the primary $T_{eff}$ for each source during the final model fitting process.

%

\begin{figure*}[!ht]
\begin{center}
\label{q_para}
\subfigure{\includegraphics[width=6cm, height=4.5cm]{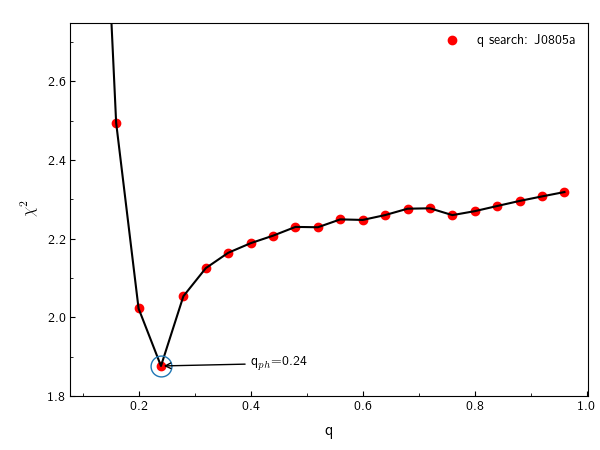}}
\subfigure{\includegraphics[width=6cm, height=4.5cm]{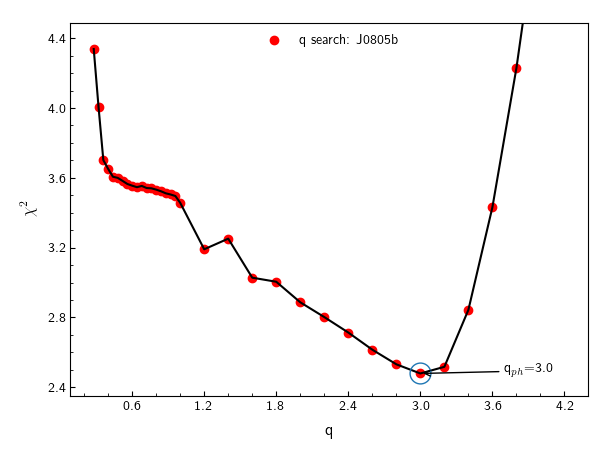}}\vspace{-0.3cm}
\subfigure{\includegraphics[width=6cm, height=4.5cm]{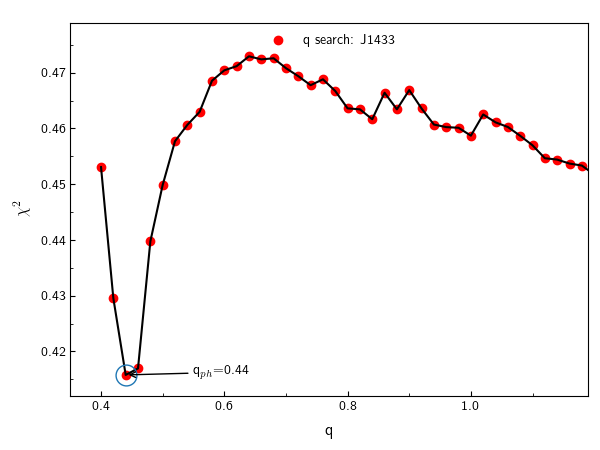}}
\subfigure{\includegraphics[width=6cm, height=4.5cm]{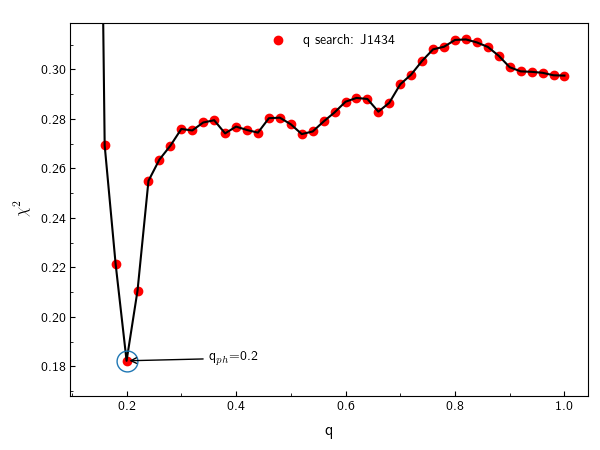}}
\caption{The $\chi^{2}$ variation with different q-parameter. The region around the lowest q is zoomed in the small subplots. }
\end{center}
\end{figure*}
\subsection{Light Curve Solution}\label{q_param}
The mass-ratio (q) is an important photometric element in the study of CB evolution as the shape of common envelop around the CB components depends on surface potential and q. In this section, we derived the photometric mass-ratio ($q_{ph}$) for each of the sources using TESS data. Generally, multi-epoch radial velocity (RV) observations are required to determine accurate spectroscopic mass-ratio ($q_{sp}$) for an EB system. However, in the absence of RV data, photometric data can also be used to determine the $q_{ph}$, provided that the EB is a total eclipsing system. \cite{2005Ap&SS.296..221T} used the synthetic data for semi-detached EBs (SDBs) and CBs to study the accuracy of $q_{ph}$. \cite{2005Ap&SS.296..221T} found that the $q_{ph}$ were more reliable for completely eclipsing SDBs and CBs as the relative stellar radii measurements were more accurate in their cases. According to \cite{2013CoSka..43...27H}, the accuracy of parameters determined using the photometric data alone increases with the amplitude of variation of CB and the precision of the Fourier coefficients. The amplitude of variation and precision of the Fourier coefficients are directly linked with the inclination angle, photometric data quality and the long period observational coverage of the system.

\begin{table}[!ht]
\caption{$T_{eff}$(in K) determined using different methods. The errors are reported in parentheses.}             
\label{all_temp}      
\centering
\scriptsize
\begin{tabular}{l l l l l}    
\hline\hline     
 Source       &    J0805a    &       J0805b & J1433        & J1434          \\
\hline
SED           & 5811(27)  & 5386(11)  & 5492(21)  & 5835(15)    \\
LAMOST        & 5608(81)  & 5238(22)  & 5687(36)  & 6014(47)    \\
GAIA          & 5578(46)  & 5629(17)  & 5507(11)  & 5682(15)    \\
T vs(J-H)    & 5397(197) & 5221(135) & 5877(151) & 6095(142)   \\
mean $T_{eff}$& 5599(55)  & 5369(35)  & 5641(39)  & 5907(38)    \\
\hline\hline               
\end{tabular}
\end{table}

The q-search method was used to determined the $q_{ph}$ for these sources using TESS photometry. The technique is quite popular in the literature and used by multiple authors time to time (e.g. \citealt{2020AJ....159..189L, 2021AJ....162...13L, 2021AJ....161..221P, 2022ApJ...927...12P, 2023ApJ...956...49L}). After loading the data in PHOEBE, TESS photometric filter and the contact binary model were selected. Updated ephemeris information was used to convert the BJD vs Flux time series to Phase vs Flux. The primary $T_{eff}$ was fixed as estimated in Section \ref{teff}. As these CBs are expected to have convective envelope, the gravity darkening coefficients ($g_{1}$ = $g_{2}$) and bolometric albedos ($A_{1}$ = $A_{2}$) were fixed to 0.32 and 0.5, respectively. In case of CBs, primary component surface potential ($\Omega_{1}$) equals secondary component surface potential ($\Omega_{2}$). The synchronicity parameters were fixed to 1 and circular orbits were assumed. The limb darkening coefficients were estimated using square root law on the basis of \cite{1993AJ....106.2096V} tables by PHOEBE itself. Out of all the parameters, we kept inclination angle (i), secondary component $T_{eff}$, $\Omega_{1}$, primary component luminosity ($L_{1}$) as adjustable parameters during iteration. For determining the $q_{ph}$, multiple models were generated for each system with q varying from 0.1 to higher values. For models with q<1, the step size was kept 0.02 and for higher q the step size of 0.25 was used (models with q=0.10, 0.12, 0.14, ..., 1.0, 1.25, 1.50 and so on). With the help of differential correction minimization method in PHOEBE, we tried to minimize the $\chi^{2}$ for each of these models for every source. Using the PHOEBE-scripter, we ran the minimization process multiple times and also introduced a random shift in the parameters from time to time. The random shift in the parameters helps the solution to escape from local minima and detect the global minima. A model with minimum $\chi^{2}$ was considered the best and q corresponding to that model was taken as $q_{ph}$. The Figure~\ref{q_para} shows the variation of $\chi^{2}$ for different models. For systems J0805a, J0805b, J1433, and J1434, the q-search estimated $q_{ph}$ as 0.24, 3.0, 0.44, and 0.2, respectively.

\textbf{J0805a: }After getting the rough estimates of $q_{ph}$ from the q-search method, q was also selected as an adjustable parameter. Following multiple iterations, q for J0805a, converged to 0.247 (0.001). The $T_{eff}$ for the secondary component was found to be 5692 (8) and the inclination angle for the system was determined as $82.1^{\circ}$ (0.1). The secondary is hotter than the primary component with a $T_{eff}$ difference of 93 K. \cite{1970VA.....12..217B} categorized EWs into two subtypes, W and A, using RV observations. For W-subtypes, the primary minima results from occultation, whereas for A-subtypes, it is due to transit/eclipse. A-subtypes are predominantly early-type stars, whereas W-subtypes are typically late-type stars. The spectral type of J0805a is listed as G3 in the LAMOST database. Although RV data are essential for accurately labeling these CB subtypes, we used photometric solutions for this classification. As the primary component of the system is more massive, larger, and is eclipsed during the primary minima, we classify this as an A-subtype system. This classification could be validated via RV measurements if the system is observed in the future by LAMOST MRS. The fill-out factor (f) of a contact binary is a good measure of degree of contact. It varies from -1 to 0 for detached systems while goes 0 to 1 for contact systems. A value closer to 1 indicates high degree of contact between binary components. The fill-out factor is calculated using the following relation:

\begin{equation}
\label{fil_CB}
\begin{aligned}
f=\dfrac{(\Omega_{inner} - \Omega_{1})}{(\Omega_{inner} - \Omega_{outer})}
\end{aligned}
\end{equation} 
Here, $\Omega_{inner}$ and $\Omega_{outer}$ are the surface potentials at the inner and outer Lagrange points, respectively. $\Omega_{1}$ represents the primary component surface potential. The $\Omega_{inner}$ is slightly higher than $\Omega_{1}$, bringing it very close to the characteristics of contact binary systems. The fill-out factor (f) is calculated as 0.5\% for the system. 

%
\begin{figure*}[!ht]
\begin{center}
\label{J0805_fit}
\subfigure{\includegraphics[width=14cm, height=4.5cm]{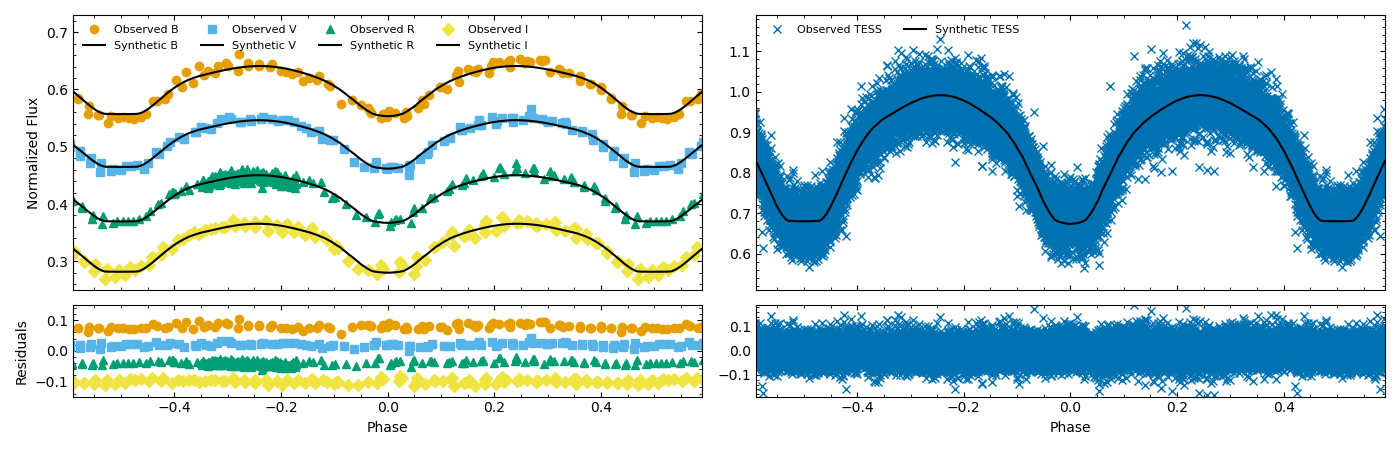}}\vspace{-0.0cm}
\subfigure{\includegraphics[width=14cm, height=4.5cm]{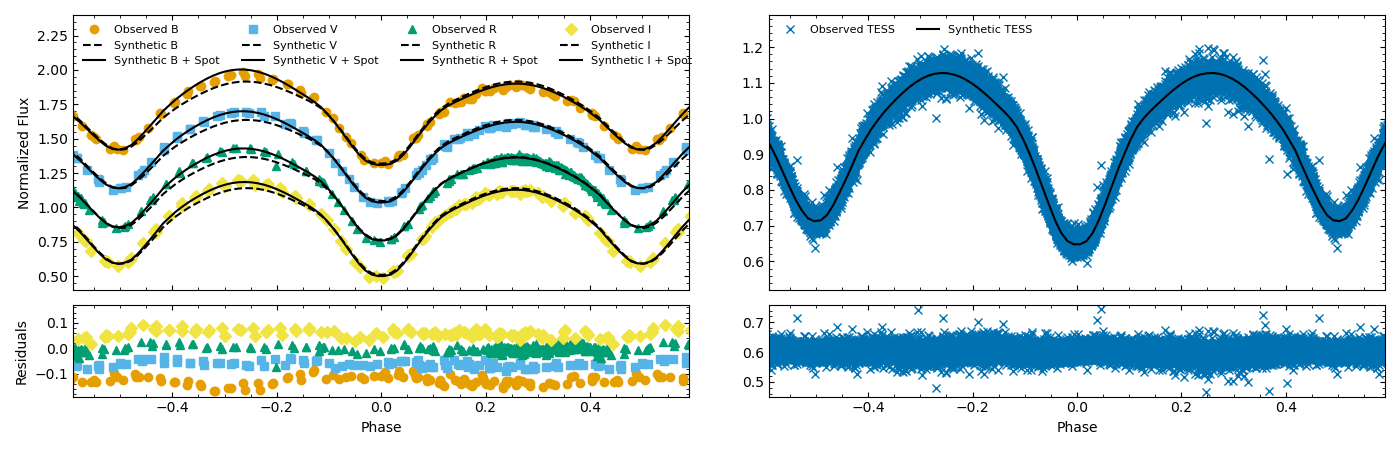}}
\caption{ The synthetic LCs (without spot: black dashed line, with spot: black continuous lines) are plotted along with observed data for J0805a (upper plot) and J0805b (lower plot).}
\end{center}
\end{figure*}

\textbf{J0805b: }In case of J0805b, $q_{ph}$ from q-search was 3.0 but converged to 3.01 (0.01) when q was also varied along with other parameters during final iterations. The $q_{ph}$, which is greater than 1 indicates that secondary is more massive than the primary component. The inclination angle (i) was calculated as $79.5^{\circ}$ (0.1). The secondary component $T_{eff}$ was determined as 4981 K, which is 388 K less than the primary  $T_{eff}$. This system is classified as a W-subtype CB because the secondary component is massive and is occulted during the primary minima phase. The spectral class for this system is reported as G7/G8 by LAMOST survey. Using the relation~\ref{fil_CB}, the fill-out factor f for the system was calculated as 17.5\%. The TESS LC for the system was symmetrical, indicating uniformity in the surface brightness while a small LC asymmetry was observed in DFOT observations. The secondary maxima (phase=-0.25) was slightly brighter than the primary maxima (phase=+0.25) in all the photometric band LCs. The brightness level at secondary maxima was found to be $\sim 0.073 $, $\sim 0.068 $, $\sim 0.051 $, and $\sim 0.045 $ mag higher than the primary maxima brightness level in B, V, R, and I, bands respectively. Under normal conditions, both the maxima are expected to have similar brightness because same amount of cross-section area is visible to the observer at phases -0.25 and +0.25. The presence of one or more hot/cool spots on the stellar surface could be one possible explanation for this observed asymmetry. Other possibilities include, circumstellar dust and gas clouds, mass transferring gas streams causing high brightness regions, and asymmetric circumfluence due to Coriolis forces \citep{2009SASS...28..107W}. To include the effects of observed asymmetry of the LCs in DFOT solutions, we used a hot spot on the secondary component. The spot position defining parameters, longitude and the latitude were fixed to $300^{\circ}$ and $80^{\circ}$ after inspecting multiple combinations. The spot radius and the spot temperature relative to the component temperature were determined as $15^{\circ}$ and 1.13, respectively. The inclusion of the spot improved the quality of the fit in the DFOT data. The fitted LCs along with observed data in BVRI and TESS band are shown in Figure~\ref{J0805_fit} for J0805a and J0805b.

\begin{table}[!ht]
\caption{The LC solutions for the targets based on DFOT and TESS observations.}
\label{TDmod_para}      
\centering
\scriptsize
\begin{tabular}{l l l l l}    
\hline\hline     
Parameters               & J0805a      & J0805b      & J1433       & J1434         \\
\hline
q                        & 0.247(0.001)& 3.01(0.01)  & 0.441(0.001)& 0.198(0.002)  \\
i ($^{\circ}$)           & 81.6(0.1)   & 79.5(0.1)   & 72.8(0.1)   & 80.2(0.4)     \\
$T_{2}$                  & 5692(8)     & 4981(9)     & 5266(7)     & 5927(4)      \\
$\Omega_{1}=\Omega_{2}$  & 2.346(0.001)& 6.626(0.009)& 2.699(0.002)& 2.156(0.001)  \\
$\Omega_{in}$            & 2.347       & 6.634       & 2.727       & 2.208         \\
$\Omega_{out}$           & 2.191       & 6.015       & 2.470       & 2.086         \\

$l_{1}/l_{tot}$ (B)      & 0.765(0.001)& 0.373(0.003)&0.772(0.002) & 0.811(0.001)  \\
$l_{1}/l_{tot}$ (V)      & 0.767(0.001)& 0.352(0.002)&0.753(0.002) & 0.810(0.002)  \\
$l_{1}/l_{tot}$ (R)      & 0.771(0.001)& 0.341(0.003)&0.743(0.001) & 0.810(0.002)  \\
$l_{1}/l_{tot}$ (I)      & 0.772(0.001)& 0.327(0.002)&0.735(0.002) & 0.809(0.001)  \\
$l_{1}/l_{tot}$ (T)      & 0.773(0.001)& 0.330(0.002)&0.734(0.001) & 0.810(0.001)  \\

$r_{1}$/a                & 0.504(0.026)& 0.297(0.022)& 0.462(0.024)& 0.544(0.031)  \\
$r_{2}$/a                & 0.265(0.018)& 0.487(0.026)& 0.313(0.021)& 0.266(0.026)  \\
$f$                      & 0.5$\%$     & 17$\%$      & 11$\%$      & 42$\%$        \\
\hline\hline               
\end{tabular}
\end{table}

%
\begin{figure*}[!ht]
\begin{center}
\label{J1433_fit}
\subfigure{\includegraphics[width=14cm, height=4.5cm]{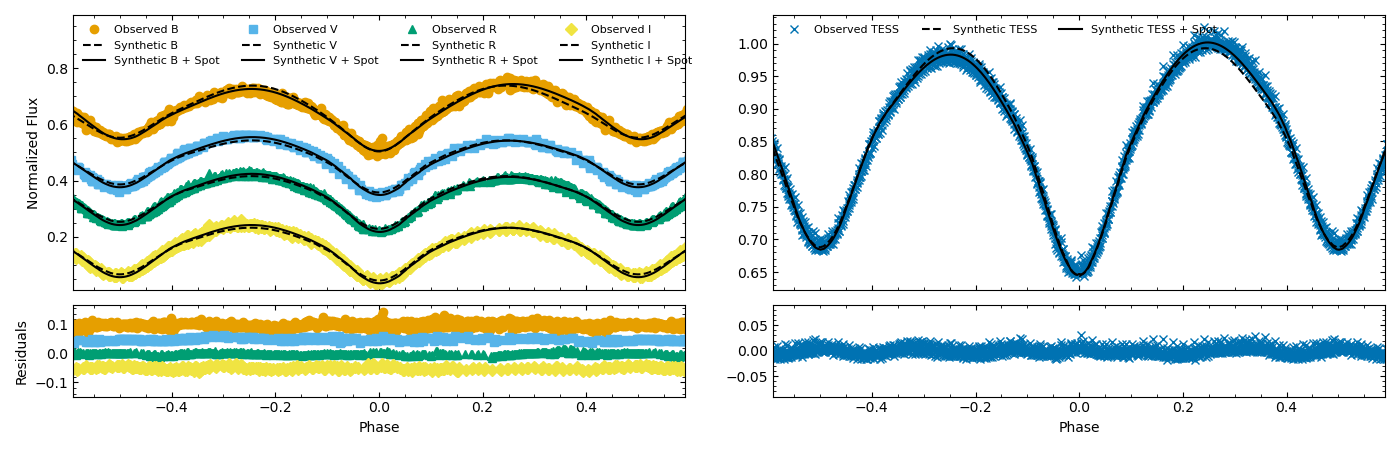}}\vspace{-0.0cm}
\subfigure{\includegraphics[width=14cm, height=4.5cm]{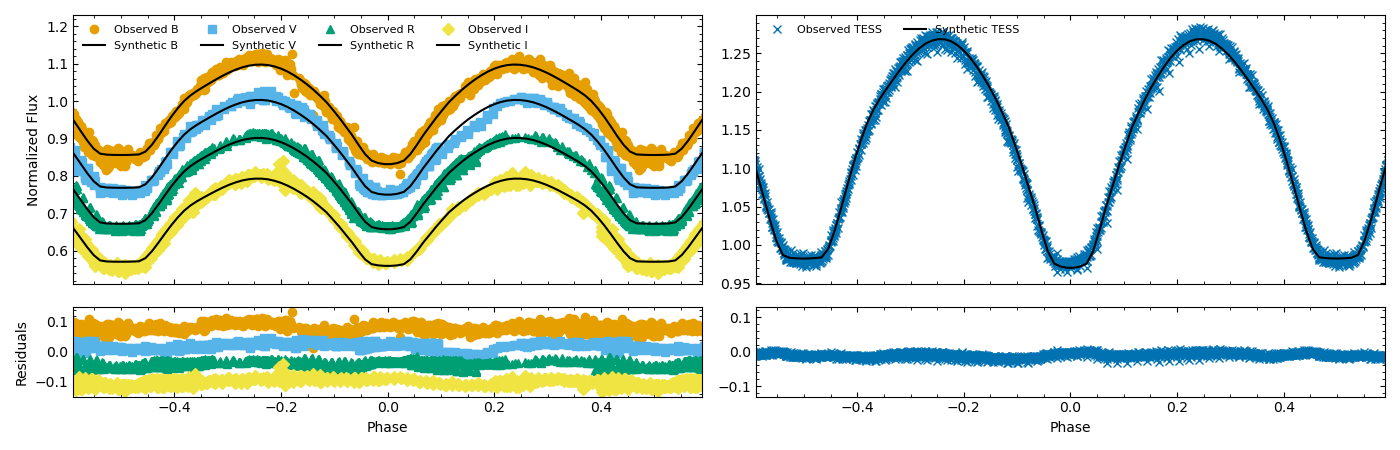}}
\caption{ The synthetic LCs (without spot: black dashed line, with spot: black continuous lines) are plotted along with the observed data for J1433 (upper plot) and J1434 (lower plot).}
\end{center}
\end{figure*}
%
\textbf{J1433: }The $q_{ph}$ was improved for J1433 by making it variable during the iteration process but did not change much from 0.44. The secondary $T_{eff}$ and the inclination angle were determined as 5266 (7) K and $71.8^{\circ}$ (0.1) for J1433 using the TESS and DFOT observations. The secondary component in the system was found to be less massive than the primary component, hence J1433 was categorized as an A-subtype CB. It has a spectral classification of G3. The primary $T_{eff}$ was 375 K higher than the secondary $T_{eff}$. The fill-out factor of the system was calculated as 11\%. A small asymmetry was observed in TESS LC as the secondary maxima was fainter than the primary maxima. Like J0805b, we used the star-spot approach to justify this asymmetry in the LC. We added a hot spot on primary component and a cool spot on secondary for TESS LC asymmetry. The parameters for the primary component hot spot were calculated as longitude = $220^{\circ}$ (fixed), latitude = $72^{\circ}$ (fixed), radius = $10^{\circ}$, and relative temperature = 1.15. Similarly, for the cool spot on the secondary component parameters were calculated as longitude = $225^{\circ}$ (fixed), latitude = $72^{\circ}$ (fixed), radius = $15^{\circ}$, and relative temperature = 0.81. In case of DFOT observations, B and VRI observations were collected at different times. The VRI band observations spread $\sim$ 680 days with most of data being collected during initial 60 days period. The B-band coverage started $\sim$ 300 days after the most of VRI observations were complete. Therefore, we observed different levels of asymmetry between B and VRI -band LCs. In case of B-band data, primary maxima was found to be 0.05 magnitude brighter than the secondary. Two hot spots, one on the secondary component and another near the equator of the primary component were used. The hot spot parameters for the primary (secondary) component were determined as longitude = $225^{\circ}$ ($315^{\circ}$) (fixed), latitude = $72^{\circ}$ ($90^{\circ}$) (fixed), radius = $15^{\circ}$ ($10^{\circ}$), and relative temperature = 1.10 (1.13), for the B-band LC asymmetry. While for VRI-band observations, a cool spot was placed on the primary component of the system as secondary maxima was slightly brighter than the primary maxima. The spot parameters were determined as longitude = $305^{\circ}$ (fixed), latitude = $75^{\circ}$ (fixed), radius = $15^{\circ}$, and relative temperature = 0.88.

\begin{figure*}[!ht]
\begin{center}
\label{spot_dis}
\subfigure{\includegraphics[width=5.8cm, height=4.35cm]{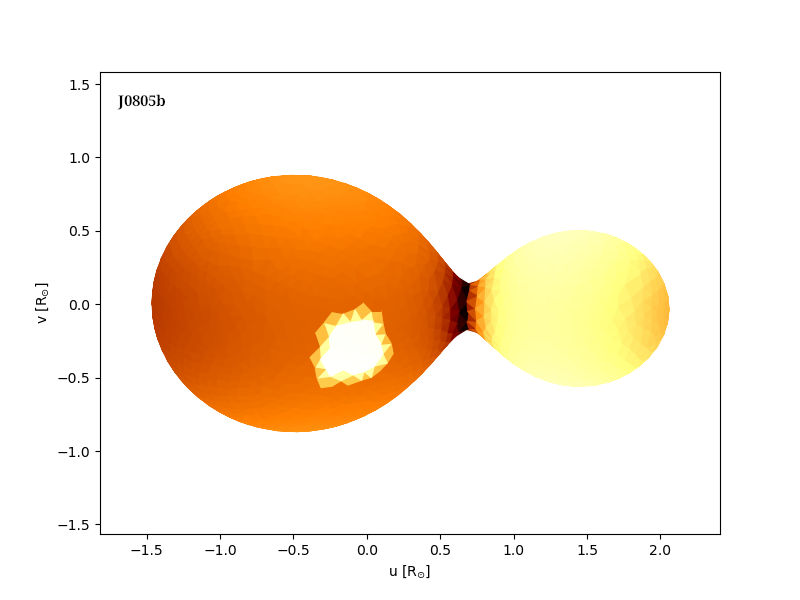}}
\subfigure{\includegraphics[width=5.8cm, height=4.35cm]{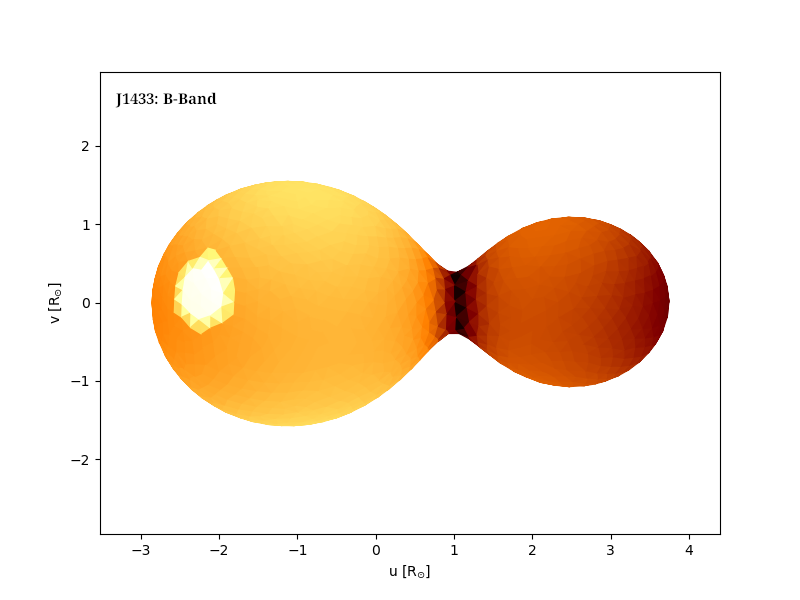}}
\subfigure{\includegraphics[width=5.8cm, height=4.35cm]{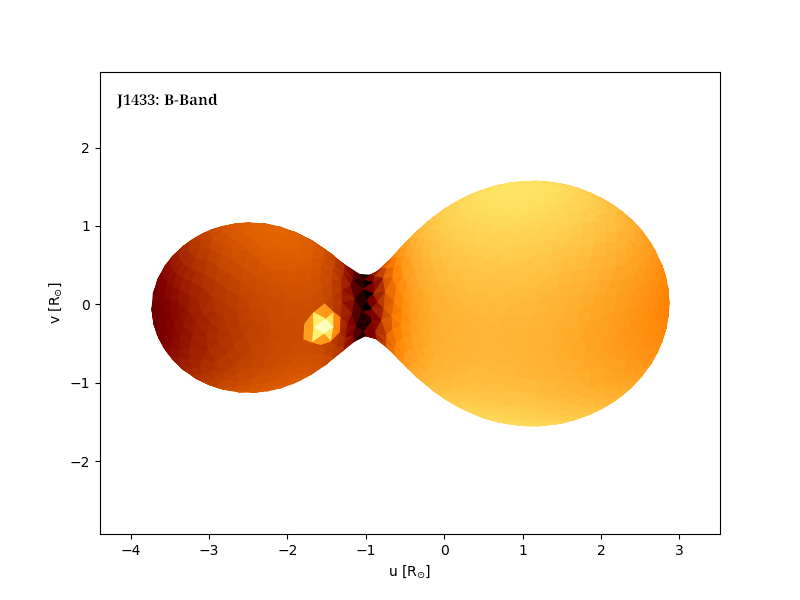}}
\subfigure{\includegraphics[width=5.8cm, height=4.35cm]{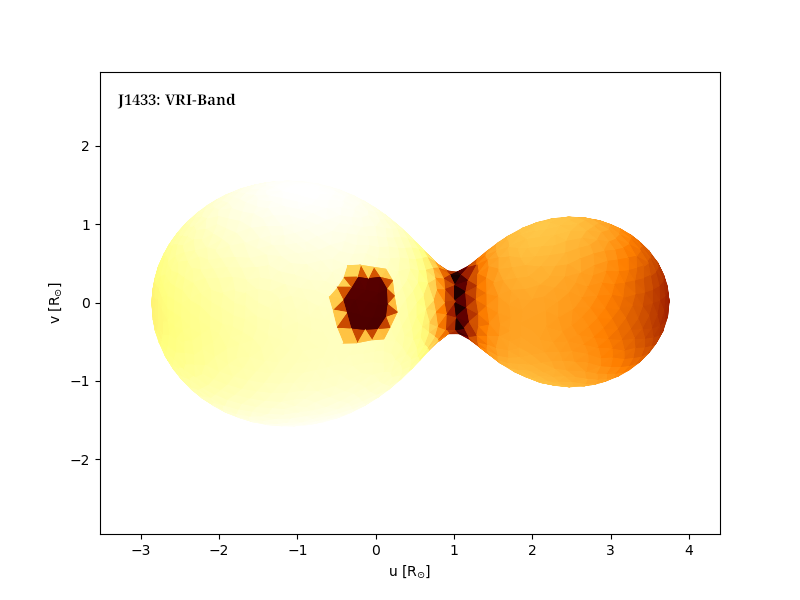}}\vspace{-0.0cm}
\subfigure{\includegraphics[width=5.8cm, height=4.35cm]{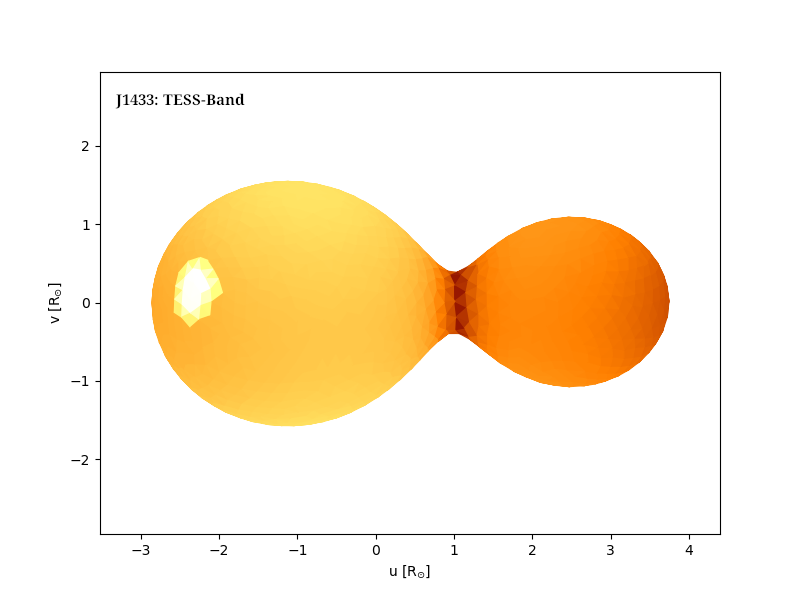}}
\subfigure{\includegraphics[width=5.8cm, height=4.35cm]{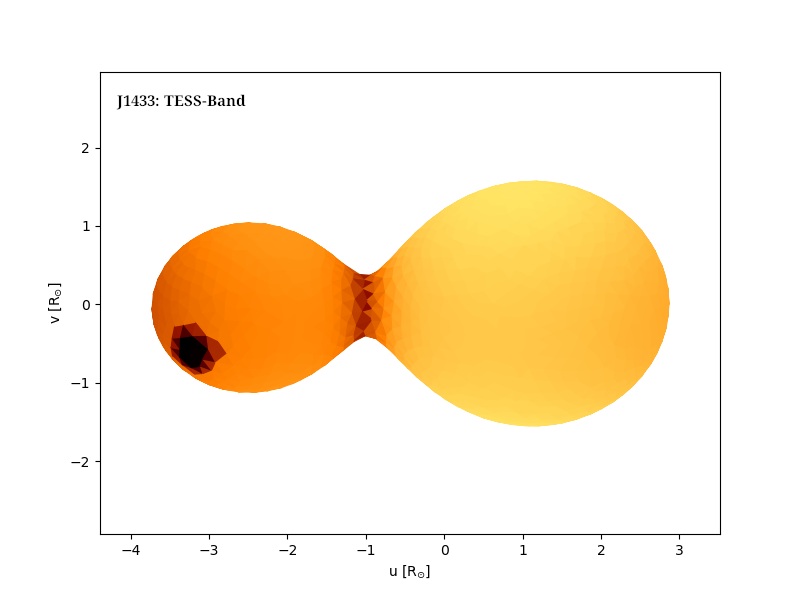}}
\caption{ The spots distributions used to explain the LC asymmetry among TESS/DFOT data are shown with red (cool spot) and blue (hot spot) dots on the stellar surface.}
\end{center}
\end{figure*}

\textbf{J1434: }For J1434, the final $q_{ph}$ was determined as 0.19 (0.01). The secondary component was $\sim$ 20 K hotter than the primary. It was found to be a partial eclipsing system with an inclination angle of $80.2^{\circ}$ (0.4). The photometric solutions suggest that J1434 is an A-subtype CB. According to the LAMOST database, its spectral classification is F9. The fill-out factor for the system was calculated as 42\%. As in the case of J1433, the J1434 observations for TESS, and VRI-band were collected at different time. A small level of asymmetry was present in all the LCs but due to the small level of asymmetry $\sim$0.01 mag, we did not use any spot on the surface of the components of the system J1434. The synthetic LCs generated using the TESS and DFOT observations are shown in the Figure~\ref{J0805_fit} and ~\ref{J1433_fit}. The LC solutions achieved via LC modeling are given in Table~\ref{TDmod_para}. The spots on CB components used to get a better fit are shown in Figure~\ref{spot_dis}.

%
\section{Physical Parameters}\label{phy_para}
In the absence of RV data, it is very hard to accurately calculate the parameters for the system and the individual components. Use of empirical relations based on a small sample of CBs is very common in the literature. \cite{2003MNRAS.342.1260Q} studied a sample of 78 CBs and reported a $M_{1}$-$P_{orb}$ relation. \cite{2006MNRAS.368.1319R} derived the M$_{v}$ -  $log \, P_{orb}$ relation using a sample of 21 CBs with $log \, P_{orb}$ < -0.25. \cite{2008MNRAS.390.1577G} performed statistical analysis on a sample 112 cool CBs and derived multiple parameter correlations including M$_{1}$- $P_{orb}$, M-R, $log (P_{orb})$-R, etc. \cite{2021ApJS..254...10L} analyzed a sample of 700 CBs compiled from 450 publications. Although RV observations were not available for most of these sample candidates, \cite{2021ApJS..254...10L} studied the masses, radii, and luminosities for the components of stars in the sample as functions of $P_{orb}$. $P_{orb}$-M and $P_{orb}$-primary $T_{eff}$ relations were derived by \cite{2022MNRAS.510.5315P} using a sample 118 CBs. \cite{2022MNRAS.510.5315P} derived the system parameters with the help of available photometric data and GAIA DR3 parallaxes. Many other empirical relations are available in the literature but these can be highly dependent on the sample size and CB targets used in the studied sample. The procedure used in this work to determine the absolute parameters involves following  steps:

\begin{table}[!ht]
\caption{The absolute parameters for the systems studied in this work.}             
\label{abs_para}
\scriptsize
\centering          
\begin{tabular}{l l l l l}
\hline\hline     
Parameters               & J0805a      & J0805b       & J1433        & J1434        \\
\hline
a($R_{\odot}$)           & 2.41(0.14)  & 2.25(0.02)   & 3.08(0.06)   & 2.73(0.06)   \\
$M_{1}$ ($M_{\odot}$)    & 1.45(0.25)  & 0.61(0.02)   & 2.19(0.15)   & 1.67(0.12)   \\
$M_{2}$ ($M_{\odot}$)    & 0.36(0.06)  & 1.86(0.06)   & 0.97(0.06)   & 0.33(0.02)   \\
$R_{1}$ ($R_{\odot}$)    & 1.21(0.07)  & 0.66(0.01)   & 1.42(0.03)   & 1.48(0.03)   \\
$R_{2}$ ($R_{\odot}$)    & 0.64(0.04)  & 1.10(0.01)   & 0.96(0.02)   & 0.73(0.01)   \\
$L_{1}$ ($L_{\odot}$)    & 1.30(0.18)  & 0.32(0.01)   & 1.86(0.09)   & 2.944(0.10)   \\
$L_{2}$ ($L_{\odot}$)    & 0.39(0.05)  & 0.68(0.01)   & 0.64(0.03)   & 0.6570.02)   \\
\hline\hline                 
\end{tabular}
\end{table}

1.) The apparent V-band magnitude ($m_{v}$) at the LC maxima was converted to the absolute V-band magnitude ($M_{v}$) of the system with the help of GAIA parallax and V-band extinction.

2.) The $M_{v}$ of the individual components were determined from the system $M_{v}$ by taking into account the $L_{1,2}/L_{tot}$ from LC modeling. Depending upon the $T_{eff}$ of the individual components, a bolometric correction (BC) was applied to the $M_{v}$ of each component to get the bolometric magnitude ($M_{bol}$). The $M_{bol}$ of each component was converted to the individual components luminosity (L).

3.) The stellar radius was calculated using the Luminosity-Radius-Temperature relation. Using the Kepler's third law, total mass of the system was determined. The individual component masses were calculated with the help of $q_{ph}$. The detailed description about the procedure is available in \cite{2020Ap&SS.365...71L, 2022ApJ...927...12P, 2024NewA..11002227P}. The absolute parameters for all of the four studied CBs are given in Table~\ref{abs_para}. The errors determined in these quantities are given in parentheses.

\section{Mass Transfer Rate}\label{ma_tr}

The period investigation on these four systems detected period change in J0805b and J1433. Three main reasons for period change in binaries are gravitational waves, magnetic braking angular momentum loss (AML), and mass-transfer between components. On the basis of J0805b component parameters, the possible period change rate due to gravitational waves is calculated as -7.81 $\times10^{-16}$ days per year. Similarly, the expected contribution in the period change rate due to magnetic braking AML is found to be -5.42 $\times10^{-8}$ days per year. Both of these values are well below than the observed period change rate in system J0805b. Therefore, the observed period change can be associated with the mass-transfer between J0805b components. The mass-transfer rate (dM$_{1}$/dt) required to explain this period change for J0805b is calculated as -1.56($\pm$0.07) $\times10^{-6} M_{\odot}$ per year. Similarly for J1433, the expected period change rate due to gravitational waves and magnetic braking AML are determined as -7.56 $\times10^{-16}$ days per year and -8.42 $\times10^{-8}$ days per year. The period change rate contribution in J1433 due to gravitational waves and magnetic braking AML does not seem to explain the observed period change rate. Therefore we can say this is also due to mass-transfer between components with a mass-transfer rate of -7.95($\pm$0.87) $\times10^{-7} M_{\odot}$ per year. The negative mass-transfer rate indicates that the mass is being transferred from primary to secondary component.

\begin{figure*}[!ht]
\begin{center}
\subfigure{\includegraphics[width=16cm,height=6.5cm]{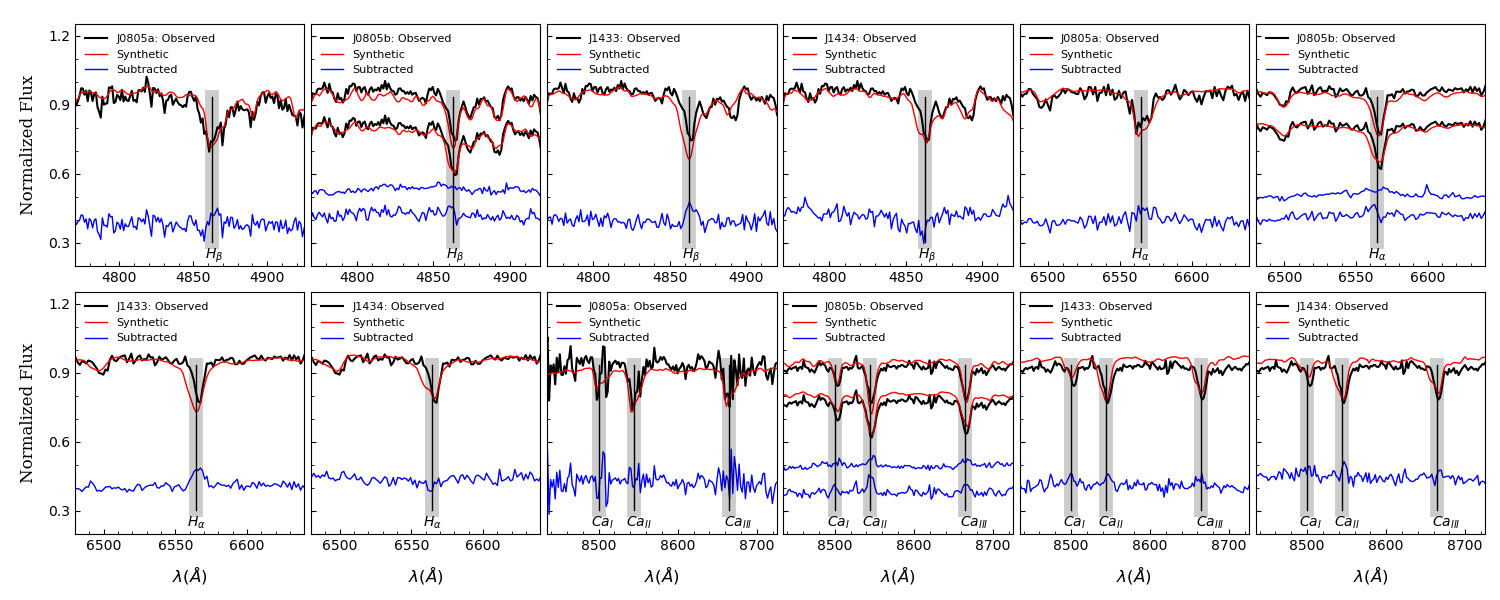}}
\caption{The $H_{\alpha}$, $H_{\beta}$, and Ca-triplet region of low resolution LAMOST spectra (black line) and synthetic spectra (red line) for the targets are shown. The subtracted spectra in the same region are shown by a continuous blue line.}
\end{center}
\label{spec_sub}
\end{figure*}

%

\section{Chromospheric Activities}\label{ch_ac}

Most of the stars from spectral class F to L and in-between evolutionary stages pre-main sequence to giant show magnetic activity phenomena. This generally happens due to the presence of thick convective zones and rapid rotation among these late type stars. The effects of magnetic activity phenomena are observable in the form of spot on stellar surface, plage, flare, long term magnetic activity cycle, etc. As CBs consist of a pair of late type stars and have high rotational velocity, they are expected to exhibit these phenomena.

Photometry and spectroscopy are the most common ways to study the magnetic activity in late type stars. Photometry helps to detect star-spots and determine their parameters like position, size and the temperature of star-spots. The most common techniques includes synthetic LC generation \citep{1971ApJ...166..605W, 1987ApJ...319..827B, 2022Galax..10....8T}, inversions of stellar LCs (to find position, differential rotation and cycles of spots; \citealt{2002A&A...394..505B}), and eclipse mapping method \citep{1997MNRAS.287..556C, 2014MNRAS.444.1721M}. The long term photometric data are also helpful to study the movement and the evolution of spots with time. New flare events can also be detected with the help of long term continuous observations. The spectroscopic study of magnetic activity phenomena involves Zeeman–Doppler Imaging and Doppler imaging. Zeeman-Doppler Imaging is used to understand the stellar magnetic field topologies profiles with the help of photospheric lines \citep{1989A&A...225..456S}. Doppler imaging tracks the spectral line asymmetries at different epochs and determines the spatial distribution of spots on the stellar surface \citep{1983ASSL..102..379V, 1987ApJ...321..496V}. One another popular method is spectral subtraction technique. Under this method, excess emission is investigated in certain spectra regions which is expected due to ongoing chromospheric activities. The important chromospheric activity indicators includes Mg II, H$_{\alpha}$, H$_{\beta}$, Ca-triplet. The active component in the binary system is observed to show an excess emission in these regions of spectra. \cite{1997A&AS..125..263M}, \cite{1998A&A...330..155M}, and \cite{2000A&AS..146..103M} studied multiple chromospherically active binaries and detected excess emission in H$_{\alpha}$, H$_{\beta}$, Mg-triplet, Na-doublet, He, Ca-doublet and Ca-triplet. The active binary components are found to show stronger emission or filled-in emission as compared to the single stars having same rotational velocities.

The LAMOST spectra of these targets were used to detect any excess emission in H$_{\alpha}$, H$_{\beta}$, and Ca-triplet region using the spectral subtraction technique. The spectral subtraction technique is discussed by \cite{1984BAAS...16..893B} and involves the subtraction of the photospheric flux from the observed flux. The level of photospheric flux is assumed to be the same for an active and inactive star. The remaining excess emission after subtraction is a measure of chromospheric activity level in the star \citep{1995A&AS..114..287M}. The spectral subtraction technique is implemented in a Fortran based program STARMOD. It uses two inactive star spectra and combine them to prepare an inactive binary template. STARMOD transforms the input inactive spectra by introducing RV shifts, rotational broadening, and weights for individual components. The final binary template from STARMOD is then subtracted from the target star spectra and residual is studied for any excess emission. Already known inactive stars with spectral type similar to our targets were searched among different catalogs (e.g. \citealt{1999ApJS..123..283M, 2000yCat..41420275S, 2004ApJS..152..251V}). One component for the synthetic spectra was selected on the basis of spectral type reported in the LAMOST catalog. For the second component, a number of inactive star spectra were tested and an appropriate star was selected on the basis of minimum $\chi^{2}$. For J0805a, HD 102800 (G7 type star) and HD 233641 (F8 type star) spectra were used to create a synthetic template. For J0805b, HD 233389 (G5 type star) and HD 246132 (G3 type star) were used to create the synthetic spectra. BD +09 1627 (G8 type star) and HD 233641 (F8 type star) spectra were used for J1433 while HD 13357 (G3 type star) and BD +39 2723 (G8 type star) were used for J1434. Figure~\ref{spec_sub} shows the H$_{\beta}$, H$_{\alpha}$, and Ca-triplet region of the LAMOST spectra along with synthetic and subtracted spectra in the same region. J0805a, J0805b, and J1434 do not show any reasonable excess emission in the studied regions. However, J1433 have shown some excess emission features in H$_{\alpha}$ and H$_{\beta}$ region. 

\begin{figure*}[!ht]
\begin{center}
\includegraphics[width=16cm,height=6cm]{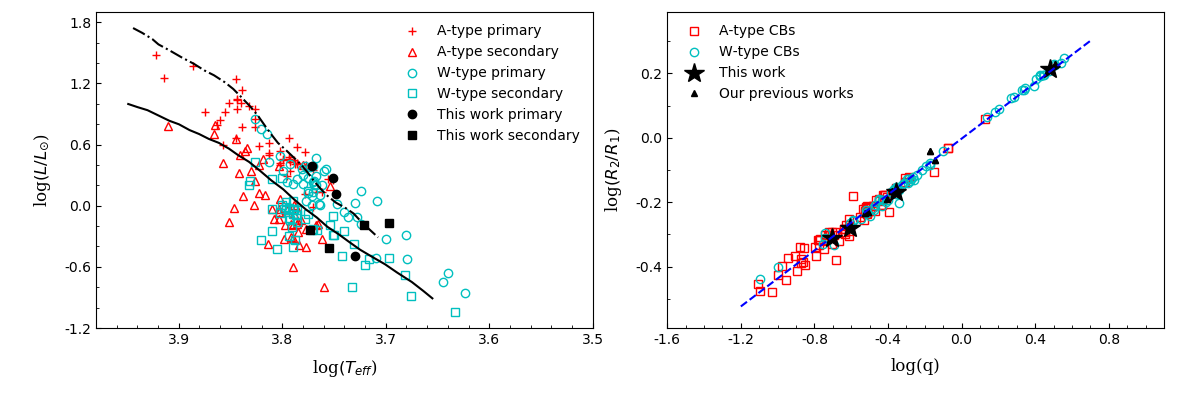}
\caption{The left plot shows the positions of the components of the system are shown in the H-R diagram with the theoretical ZAMS (continuous black line) and TAMS (dash dot black line) lines. The right plot shows the mass-ratio and the radius ratio relation for the targets including previously studied CBs.}
\label{HR_dia}
\end{center}
\end{figure*}

\begin{figure*}[!ht]
\begin{center}
\includegraphics[width=16cm,height=6cm]{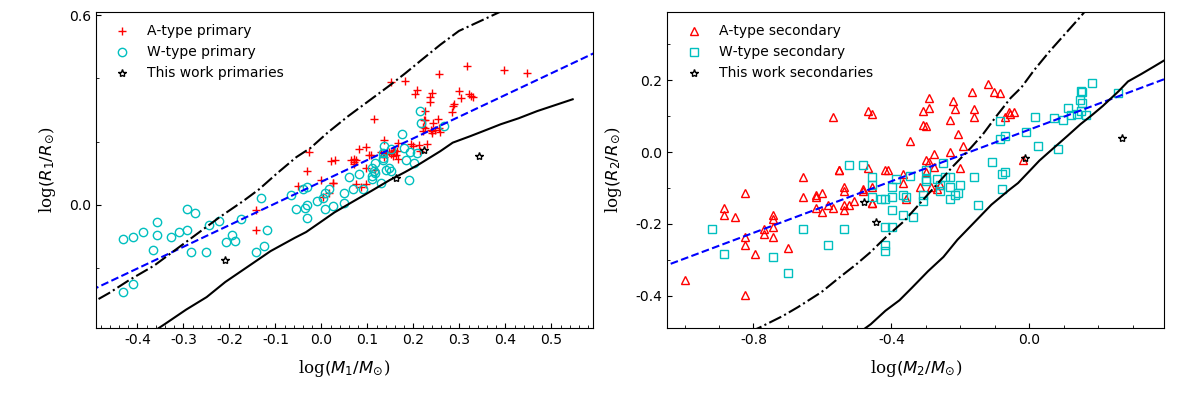}
\caption{The $M_{1}$-$R_{1}$ and $M_{2}$-$R_{2}$ diagram for the targets including previously studied CBs.}
\label{HR_rela}
\end{center}
\end{figure*}
%
\section{Results and Discussion}\label{discu}
Four CB systems from the Catalina Surveys Periodic Variable Star Catalog are studied with the help of multi-band photometric and low-resolution LAMOST spectroscopic data. The photometric data involves BVRI-band observations from DFOT and TESS observations. Based on the O-C analysis, it is concluded that systems J0805b and J1433 have variable $P_{orb}$. Possible mechanisms behind the reported period change were studied, and mass transfer was found to provide a reliable explanation for the observed period variation. The synthetic LCs are generated with the help of PHOEBE software package. All the systems were found to have q$_{ph}$ or (1/q$_{ph}$) less than 0.5. It should be noted that the q$_{ph}$ estimates are not always reliable. As discussed by \cite{2021AJ....162...13L}, the reliability of q$_{ph}$ is highly dependent on the inclination angle of the system. Among a sample of 101 CBs, \cite{2021AJ....162...13L} found that the q$_{ph}$ were very different from the q$_{sp}$ in case of partial eclipsing CBs. Except J1433 all the systems have inclination angle >79$^{\circ}$. A small asymmetry was seen in LCs for J0805b and J1433. We used cool/hot spots on stellar surface of components to introduce the contribution of observed asymmetry in the synthetic LCs. The low accuracy of observational data can create the non-uniqueness of the spot solution. Due to this, multiple combination of spot parameters can generate similar effect on the LCs. According to \cite{1999NewA....4..365E}, the reliable mapping of stellar surface requires high precision observations with sufficient coverage of rotation phases. \cite{1999NewA....4..365E} argued that the accuracy should be better than 0.1 mmag. The accuracy of DFOT data is not as good as TESS observations. Therefore the spot parameters determined from DFOT observations might not be very reliable in our analysis. Though the small excess emission in J1433 LAMOST spectra in the H$_{\alpha}$ region can be an indication of magnetic activity in this system. The TESS part of O-C diagram of J1433 also shows some short term variation possibly due to star spot evolution with time. Three systems have shallow contact configuration with fill-out factor less than 25$\%$ while J1434 has moderate contact configuration.

The physical parameters were determined with the help of GAIA DR3 parallax. This technique is expected to be more reliable, as other empirical relations may be biased toward the sample of CBs from which they are derived. Figure~\ref{HR_dia} shows the position of CB components on the HR diagram. Other well studied CBs from previous literature \citep[e.g.,][]{2011MNRAS.412.1787D, 2013MNRAS.430.2029Y} are also plotted along with investigated targets. Due to component interaction, the evolution of contact binaries is assumed to be different from isolated single stars or long period detached binary system component. However the deviation of less massive components on the HR diagram from the ZAMS line can be associated with mass/energy transfer from more massive to less massive component. The right side of the Figure~\ref{HR_dia} shows the q vs (R$_{2}$/R$_{1}$) relation for studied CBs. The relation is updated as follows:

\begin{equation}
\label{qr_rela}
\begin{aligned}
log(R_{2}/R_{1}) &= 0.434 (\pm 0.004) \times log (q)\\
&-0.003 (\pm0.002)
\end{aligned}
\end{equation}

Figure~\ref{HR_rela} shows the primary and secondary component masses relation with primary and secondary component radii. The massive components of each of the system are close to or above ZAMS while less massive components are close to or above TAMS in the M-R plot. The linear fit to the sample is shown with blue dashed line. The updated relations follows following trend:
\begin{equation}
\label{m1r1_rela}
\begin{aligned}
log(M_{1}) &= 0.686 (\pm0.028) \times log (R_{1})\\
&+0.073 (\pm 0.006)
\end{aligned}
\end{equation}
\begin{equation}
\label{m2r2_rela}
\begin{aligned}
log(M_{2}) &= 0.359 (\pm0.024) \times log (R_{2})\\
&+0.063 (\pm0.011)
\end{aligned}
\end{equation}

\cite{2020ApJS..247...50S} studied the CRTS CBs and published the physical parameters. J0805a and J0805b are also mentioned in the catalog by \cite{2020ApJS..247...50S}. The q is determined as 0.2 and 0.31 for J0805a and J0805, respectively. \cite{2020ApJS..247...50S} reported J0805b as a total eclipsing system with inclination ~85$^{\circ}$ but in our case it was around 80$^{\circ}$. As \cite{2020ApJS..247...50S} used M-L relation by \cite{2013MNRAS.430.2029Y} to derive the physical parameters of the primaries, assuming they are main sequence stars, we implemented a different approach to derive the parameters, resulting in deviations among the determined values. \cite{2024ApJS..271...32L} studied J0805b, J1433, and J1434 using the ASAS-SN observations via MCMC and neural networks. The q and inclination in \cite{2024ApJS..271...32L} for J0805b matches well with our results while for J1433 we see a big difference in reported q (=0.2). For J1434,  \cite{2024ApJS..271...32L} reported q=0.61 and i=66$^{\circ}$. \cite{2024ApJS..273...31W} investigated the CRTS CBs using neural network machine learning technique and one of the system reported was J1434. The inclination and q are reported as 76$^{\circ}$ and 0.4, which are very close to our estimates. The discrepancy in the system parameters across different studies can be attributed to the varying techniques used to derive the physical parameters. The techniques used by \cite{2024ApJS..271...32L} and \cite{2024ApJS..273...31W} involves machine learning which depends on the trained data set and use only single band data. Similarly, results by \cite{2020ApJS..247...50S} are derived using only CRTS LCs. As in our case we are using, multi-band observations including good quality TESS LCs, we believe our reported results are more reliable.

We collected low resolution LAMOST spectra for these sources to investigate the chromospheric activities in form of excess emission. We used H$_{\alpha}$, H$_{\beta}$, and Ca- triplet region of the spectra for our analysis. The asymmetry was clearly observable in the DFOT LCs for J0508b, J1433, and J1434. However, the spectra analysis showed excess emission in he case of J1433 only. The targets are observed using DFOT, TESS, and LAMOST at different times, so, we do not expect clear correlation between excess emission and LC asymmetry. The TESS observations did not exhibit any clear asymmetry for targets except J1433. This suggests that the spots may not present during the TESS observations. The spots causing the LC asymmetry can appear, migrate, and disappear with time. Even for J1433, it was impossible to determine the individual component contribution in the observed excess emission. High resolution spectroscopy can help to confirm the excess emission and measure the contribution by each of components.

\section{ACKNOWLEDGEMENTS}
Guoshoujing Telescope (the Large Sky Area Multi-Object Fibre Spectroscopic Telescope LAMOST) is a National Major Scientific Project built by the Chinese Academy of Sciences. Funding for the project has been provided by the National Development and Reform Commission. LAMOST is operated and managed by the National Astronomical Observatories, Chinese Academy of Sciences. In this work we have also used the data from the European Space Agency (ESA) mission GAIA, processed by the GAIA Data Processing and Analysis Consortium (DPAC). This work also make use of the Two Micron All Sky Survey and SIMBAD database. Work at the Physical Research Laboratory is supported by the Department of Space, Govt. of India.

\bibliographystyle{yahapj}
\bibliography{Bibilography}

\end{document}